\documentclass[12pt]{iopart}

\usepackage[utf8]{inputenc}
\usepackage[T1]{fontenc}
\usepackage{color}
\usepackage{epsfig,amsfonts}
\usepackage{graphicx}
\usepackage{float}
\usepackage{epstopdf}
\usepackage{blindtext}

\def\be{\begin{equation}}
\def\ee{\end{equation}}

\def\bea{\begin{eqnarray}}
\def\eea{\end{eqnarray}}

\def\ben{\begin{enumerate}}
\def\een{\end{enumerate}}

\def\bea{\begin{eqnarray}}
\def\eea{\end{eqnarray}}

\begin{document}

\title[Local magnetization in Richardson-Gaudin models]{ Steady-states of out-of-equlibrium inhomogeneous Richardson-Gaudin quantum integrable models in quantum optics
}
\author{Hugo Tschirhart$^{1}$, Thierry Platini$^{2}$, Alexandre Faribault$^{3}$}
\address{
$^{1}$ Physics and Materials Science Research Unit, University of Luxembourg, L-1511 Luxembourg
\\
$^{2}$ Applied Mathematics Research Center, Coventry University, Coventry, England
\\
$^{3}$  Universit\'e de Lorraine, CNRS, LPCT, F-54000 Nancy, France
}
\ead{hugo.tschirhart@uni.lu, alexandre.faribault@univ-lorraine.fr}
\begin{abstract}

In this work we present numerical results for physical quantities in the steady-state obtained after a variety of product-states initial conditions are evolved unitarily, driven by the dynamics of quantum integrable models of the rational (XXX) Richardson-Gaudin family, which includes notably Tavis-Cummings models. The problem of interest here is one where a completely inhomogeneous ensemble of two-level systems (spins-1/2) are coupled to a single bosonic mode.

The long-time averaged magnetisation along the z-axis as well as the bosonic occupation are evaluated in the diagonal ensemble by performing the complete sum over the full Hilbert space for small system sizes. These numerically exact results are independent of any particular choice of Hamiltonian and therefore describe general results valid for any member of this class of quantum integrable models built out of the same underlying conserved quantities. 

The collection of numerical results obtained can be qualitatively understood by a relaxation process for which, at infinitely strong coupling, every initial state will relax to a common state where each spin is in a maximally coherent superposition of its $\left|\uparrow\right>$ and $\left|\downarrow\right>$ states, i.e. they are in-plane polarised, and consequently the bosonic mode is also in a maximally coherent superposition of different occupation number states. This bosonic coherence being a feature of a superradiant state, we shall loosely use the term superradiant steady-state to describe it.  

 A finite value of the coupling between the spins and the bosonic mode then leads to a long-time limit steady-state whose properties are qualitatively captured by a simple "dynamical" vision in which the coupling strength $V$ plays the role of a time $t_V$ at which this "relaxation process" towards the common strong coupling superradiant steady-state is interrupted.

\end{abstract}

\date{\today}

\section{Introduction}

In recent years, the technical progress has enabled the experimental study of low dimensional quantum systems, including some which can be well described by quantum integrable models (\cite{hietarinta_classic_vs_quantum_integrability} and \cite{caux_interability} present interesting discussions about, respectively, the difficulty of defining the notion of quantum integrability compared the classical one and different possible definitions of quantum integrability). As examples, experimentalists have been able to study excitons and polaritons in quantum wells \cite{{bose-einstein_exciton_polarition1},{bose-einstein_exciton_polarition2},{bose-einstein_exciton_polarition3}} and semi-conductors quantum dots \cite{{koppens_sc_quantum_dots},{petta_sc_quantum_dots},{hanson_sc_quantum_dots},{berezovsky_sc_quantum_dots},{hennessy_sc_quantum_dots},{pioro_sc_quantum_dots},{yanwen_sc_quantum_dots}}. Additionally, a lot of experiments aim to observe the superradiant phase transition \cite{Garraway_superradiant} of the Dicke model \cite{dicke} for a Bose-Einstein condensate in optical cavities \cite{{dicke_transition1},{dicke_transition2},{dicke_cavity},{bose-einstein_multimode_cavity},{proposed_dicke_model_cavity_QED_system},{dicke_model_optical_cavity}, {dicke_hubbard_lattice},{collective_bose-einstein_optical_cavities},{keldysh_dicke_cavities}}. All these new experimental accomplishments has created a strong need for numerical work describing integrable models. This work demonstrates that mathematical objects introduced in \cite{bosonhugo} provides remarkable possibilities for the numerical treatment of a family of spin-boson integrable models of the Richardson-Gaudin class\cite{gaudin}.


In this work we use the integrability of the inhomogeneous Tavis-Cummings Hamiltonian \cite{{TC-1},{TC-2}}:

\bea
H_{TC} = \omega b^{\dag}b + \sum_{j=1}^{N}\epsilon_{j}S_{j}^{z} + V\sum_{j=1}^{N}(b^{\dag}S^{-}_{j}+S^{+}_{j}b),
\label{dickeh}
\eea
\noindent one of the many Hamiltonians which belong to the general class of spin-boson Richardson-Gaudin integrable quantum models in order to study numerically the properties of the steady-state it reaches at long times following its unitary time-evolution from given initial states. The model is characterised by (Zeeman) gaps $\epsilon_j$ which are here chosen distinct for each of the $N$ spins (two-level atoms) present. Moreover, it allows, through a common coupling strength $V$ the flipping up or down of each individual spin through the absorption or emission of a single boson of energy $\omega$. The Tavis-Cummings models can be understood as the Dicke model \cite{dicke} in the Rotating Wave Approximation (RWA) and can therefore be used to describe cavity quantum electrodynamic (QED) \cite{TC-1,TC-2}.


The numerical tools used in this work rely on the reparametrisation of the Algebraic Bethe Equations in terms of eigenvalue-based variables \cite{{baode1},{baode2},{faridet},{fari_bigspin}} which allows fast and stable computation of numerically exact eigenstates (and eigenvalues) as well as the scalar products and matrix elements required to study the time-evolution of Gaudin models. The recent generalisation of the mathematical work previously mentioned to models which include a bosonic mode \cite{bosonhugo} allows the same type of numerical calculation in the inhomogeneous Tavis-Cummings Hamiltonian. The resulting mathematical construction becomes particularly useful for highly excited systems. Indeed, the complexity of the expressions involved is not defined by the actual number of excitations $M$, as they would be using a traditional Bethe ansatz, but are actually limited by the system size, i.e. the number $N$ of spins considered. This reduces vastly the complexity of numerical work for systems of small size $N$ containing a large number of excitations (up to $M=100$ in this work).




In section \ref{bosonspin1/2}, we first introduce the Gaudin algebra which defines the Gaudin models and describe how the Algebraic Bethe Ansatz (ABA) allows one to find the eigenstates of such Hamiltonians and finally explicitly present the change of variable which reformulates the problem in a way which is particularly well suited for numerical treatments. In the next section \ref{formfactors}, the determinant expressions giving scalar products and form factors (matrix elements) which constitute the basic building blocks needed for dynamical studies, are presented. 

Finally, in section \ref{numresults}, numerical results for the steady-state local magnetisation and bosonic occupation of a system of $15$ spins-$\frac{1}{2}$ containing a given number $M$ of excitations are presented. The impact of the model's parameters: the coupling strength $V$ and the bosonic frequency $\omega$, in relation to the two-level systems' energy splittings, are investigated.  Moreover, a variety of initial states differing in their spin configuration and in the total number of excitations are studied.  We finally propose a qualitative picture which tries to take into account both the system's parameter and the specific initial state in order to characterise their impact on the physical properties of the steady-state reached by these unitarily evolved quantum systems.

\section{XXX Richardson-Gaudin models containing $N$ spins-$\frac{1}{2}$ and a single bosonic mode}
\label{bosonspin1/2}

The XXX Richardson-Gaudin models contains a single bosonic mode as well as $N$ spins-$\frac{1}{2}$ and are all based on the following realisation of the generalised Gaudin algebra, given by an infinite set of generators parametrised by a spectral parameter $u \in \mathbb{C}$:

\bea
\mathrm{S}^+(u) = b^\dag + \sum_{j=1}^N \frac{V}{u-\epsilon_j} S^+_j \ ,\  \mathrm{S}^-(u) = b + \sum_{j=1}^N \frac{V}{u-\epsilon_j} S^-_j  ,\nonumber\\  \mathrm{S}^z(u) = \frac{\omega-u}{2V} - \sum_{j=1}^N \frac{V}{u-\epsilon_j} S^z_j.
\label{sbreal}
\eea

Here, $b^\dag$ and $b$ obey canonical bosonic commutation rules while $S^+_i, S^-_i,S^z_i$ form $N$ copies of the well-known spin-1/2 representation of $SU(2)$. The construction $S^2(u) \equiv S^z(u)S^z(u)+\frac{1}{2} S^+(u)S^-(u)+\frac{1}{2} S^-(u)S^+(u)$ form an infinite set of operators which all commute with one another, $\left[S^2(u),S^2(v)\right] = 0 \ \forall u,v \in \mathbb{C}$, and therefore share a common eigenbasis which is found using the ABA. 

A set of $N$ "local" conserved charges $R_1, R_2 \dots R_N$ is found by looking at the operator-valued residues of $\mathrm{S}^2(u)$: 

\bea
\mathrm{S}^2(u) = \sum_{i=1}^N \frac{R_i}{u-\epsilon_i} + \left[b^\dag b + \sum_{i=1}^N S^z_i \right] + \frac{1}{2} + \left(\frac{\omega-u}{2V}\right)^2+\frac{3}{4} \sum_{i=1}^N \frac{V^2}{(u-\epsilon_i)^2},
\nonumber\\
\label{generating}
\eea

\noindent where the $\frac{3}{4}$ factors are simply the Casimir invariant of each local spin-$\frac{1}{2}$, and are given by:

\bea
R_i =  \left(\epsilon_i - \omega\right) S^z_i+V\left(b^\dag S^-_i+bS^+_i\right) + \sum_{j\ne i}^N \frac{2V^2}{\epsilon_i-\epsilon_j} \vec{S}_i\cdot \vec{S}_j.
\label{consboson}
\eea

These conserved charges all commute together and are actually supplemented by one more conserved operator (found in the $u$ independent term in $S^2(u)$), namely $\hat{M} \equiv b^\dag b + \sum_{i=1}^N \left(S^z_i+\frac{1}{2}\right)$.  Indeed, $\hat{M}$ also commutes with every $R_i$ so that the common eigenstates of the $R_i$ all have a fixed number of excitations $M$: the eigenvalue of $\hat{M}$. This number simply counts the total number of up-pointing spins plus the total number of bosonic excitations and therefore defines a total number of excitations in the system.

Any Hamiltonian constructed as an arbitrary linear combination of $\hat{M}$ and the $N$ conserved charges $R_i$, i.e. $H = \gamma \hat{M} + \sum_{i=1}^N \alpha_i R_iÊ$, will share the same set of eigenstates. This includes the inhomogeneous Tavis-Cummings model (\ref{dickeh}) which can be written as $H_{TC} = \omega \hat{M} + \sum_{i=1}^N R_i$. The sum over $i$ cancels any explicit spin-spin coupling in the Hamiltonian and therefore, any spin-spin interaction has to be mediated by the common interaction with the bosonic mode, with which each spin has direct interaction. On the other hand, one can also look at the specific Hamiltonian associated with a single one of these conserved charges, as $H \equiv R_1$ for instance. This Hamiltonian describes a central spin $S_1$ which is coupled through hyperfine contact interaction with a bath of non-interacting spins as well as with the bosonic mode. In this case, a single spin is in contact with the bosonic mode, and each other spin is only in direct interaction with that single central spin. Despite their apparently radically different physics, both of these models have the same set of common eigenstates, all of them found through the ABA.

These eigenstates are found by first constructing a generic state through the repeated application of $\mathrm{S}^+(u) = b^\dag + \sum_{j=1}^N \frac{V}{u-\epsilon_j} S^+_j$  above a pseudo-vacuum defined here by a fully down-polarised spin sector and an empty bosonic mode:
\bea
\left|\lambda_1 \dots \lambda_M\right> = \prod_{i=1}^M \mathrm{S}^+(\lambda_i) \left|0; \downarrow, \dots \downarrow\right>.
\label{bethelikeboson}
\eea

As it should be for the ABA to apply, this vacuum $\left|0; \downarrow, \dots \downarrow\right>$ is an eigenstate of both $\mathrm{S}^2(u)$ and $\mathrm{S}^z(u)$ and is annihilated by any $\mathrm{S}^-(u)$. The direct application of $\mathrm{S}^2(u)$ on the generic previous state (\ref{bethelikeboson}) shows that this generic constructions becomes one of the eigenstates of $\mathrm{S}^2(u)$ whenever the Bethe roots $\{\lambda_1 \dots \lambda_M\}$ form a solution to the algebraic system of Bethe equations:
\bea
\frac{\omega-\lambda_i}{2V^2} + \frac{1}{2} \sum_{k=1}^N \frac{1}{\lambda_i - \epsilon_k} = \sum_{j\ne i}^M \frac{1}{\lambda_i - \lambda_j}.
\label{bethebosrap}
\eea

Any solution to this set of $M$ algebraic equations then defines one given eigenstate whose eigenvalues of the conserved charges (\ref{consboson}) are respectively given by:

\bea
r_i = \frac{V^2}{2} \left(\sum_{j\ne i}^N \frac{1}{\epsilon_i-\epsilon_j}\right) -\frac{\epsilon_i-\omega}{2} - V^2 \sum_{j=1}^M \frac{1}{\epsilon_i - \lambda_j}.
\label{ripart}
\eea 

Having a fixed number $M$ of excitations it is t also an eigenstate of the $\hat{M}$ operator with eigenvalue $M$. The specification of the set $\{\lambda_1 \dots \lambda_M\}$ and its cardinality $M$ completely defines the corresponding eigenstate, but through a change of variables, it is also possible to build an equivalent description in terms of the $N$ variables $\Lambda(\epsilon_i) =  \displaystyle \sum_{j=1}^M \frac{1}{\epsilon_i - \lambda_j}$ which correspond to the non-trivial state dependent part of the eigenvalues (\ref{ripart}). In terms of these last variables, it can be shown that, for any $\omega$ and $V$, eigenstates are defined by sets $\{ \Lambda(\epsilon_1)\dots \Lambda(\epsilon_N)\}$ which are solution to the eigenvalue-based quadratic Bethe equations \cite{baode1} given, by:

\bea
 \left[ \Lambda(\epsilon_i)\right]^2 = \sum_{j \ne i}^N \frac{ \Lambda(\epsilon_i)- \Lambda(\epsilon_j)}{\epsilon_i-\epsilon_j}  -\frac{\epsilon_i - \omega}{V^2} \Lambda(\epsilon_i)+ \frac{M}{V^2}.
\label{quadbos}
\eea

These quadratic equations are much simpler to handle numerically speaking than the traditional ones  (\ref{bethebosrap})  \cite{slavnov,links,izergin,korepin} so that defining eigenstates in terms of 
$\{ \Lambda(\epsilon_1)\dots \Lambda(\epsilon_N)\}$ makes it possible to find them individually in a very efficient way. In order to fully exploit this representation, it becomes essential to express the needed quantities, such as scalar products and matrix elements of local operators, directly in terms of these state-defining variables $\Lambda(\epsilon_i)$. These recently built expressions are precisely what the next section describes.


\section{Form Factors}
\label{formfactors}

It was shown in \cite{bosonhugo} that the projection of a state of the form (\ref{bethelikeboson}) onto a tensor product state containing $M-m$ bosons with the $m$ spins of index $\{i_1\dots i_m\}$ pointing up while the $N-m$ other spins point down, can be written as an $m\times m$ determinant whose matrix elements depend exclusively on the $\Lambda(\epsilon_i)$:

\bea
\left< M-m;\uparrow_{\{i_1\dots i_m\}}\right.\left|\lambda_1 ... \lambda_{M}\right> = \sqrt{(M-m)!}\ V^{m}\ \mathrm{Det} J
\nonumber\\
\nonumber\\
J_{ab} =
 \left\{ \begin{array}{cc}
\displaystyle\sum_{c=1 (\ne a)}^m  \frac{1}{\epsilon_{i_a} -\epsilon_{i_c} }-\Lambda(\epsilon_{i_a}) & a= b
\\
\frac{1}{\epsilon_{i_a} -\epsilon_{i_b} }
& a\ne b \end{array}\right. .
\label{detrep}
\eea
\label{basicsofz}

The squared norm $N_\Lambda$ of the state $\left|\lambda_1 ... \lambda_{M}\right>$ was also shown in \cite{bosonhugo} to be computable as a $N \times N$ determinant of the same form and was later found to be actually deducible without any aditionnal calculations \cite{{pieterdeformdicke},{pieterXXZ},{pieterunifiedalgebra}}.



On the other hand, for any eigenstate, the quantum expectation values of local operators were also proven \cite{bosonhugo} to be be obtainable from the Hellmann-Feynman theorem, giving the following expression for the local magnetisation of spin $k$:
\bea
\frac{\left<\lambda_1 ... \lambda_{M}\right|S^z_k\left|\lambda_1 ... \lambda_{M}\right>}{N_{\Lambda}}&=&\left[ -\frac{1}{2} + \frac{\partial \Lambda(\epsilon_k)}{\partial \omega}\right] ,
\eea
while the bosonic occupation follows from the fact that $\hat{M}=b^\dag b + \sum_{i=1}^N \left(S^z_i+\frac{1}{2}\right)$ has a well defined eigenvalue so that for every state we trivially find:
\bea
\frac{\left<\lambda_1 ... \lambda_{M}\right|b^\dag b\left|\lambda_1 ... \lambda_{M}\right> }{N_{\Lambda}} &=&
M- \sum_{i=1}^N\left[ \frac{\partial \Lambda(\epsilon_i)}{\partial \omega}\right].\nonumber\\
\eea

Decomposing the time-evolving state $e^{-iHt}\left|\psi_{t=0}\right>$ onto the basis of eigenstates, the time dependent expectation value of any observable $O$ is generically given by a double sum over the full set of normalised eigenstates $\left| n \right>$ (with eigenvalue $E_n$):

\bea
\left< \hat{O} \right>(t) = \sum_{n,m} \left< \psi_{0}\right|\left. n\right>\left< m\right|\left. \psi_{0}\right> e^{i(E_{n}-E_{m})t}\left< n \right| \hat{O} \left| m \right>.
\eea

Although any possible Hamiltonian $H = \gamma \hat{M} + \sum_{i=1}^N \alpha_i R_i$ will share the same eigenstates, the full time evolution of any given observable would still depend on the particular choice of $H$ since its spectrum of eigenvalues $E_n = \gamma M + \sum_{i=1}^N \alpha_i r_i$ does depend on a specific choice.

However, when computing the long time average, leading to the averaging out of any non-zero frequency, only diagonal contributions ($n=m$) survive:

\bea
\lim_{t \rightarrow \infty}\frac{1}{t}\int_{0}^{t}\left< \hat{O} \right>(t')dt' =  \sum_{n} \left< \psi_{0}\right|\left. n\right>\left< n\right|\left. \psi_{0}\right> \left< n \right| \hat{O} \left| n \right>.
\eea

\noindent The resulting diagonal ensemble no longer involves any eigenvalues and therefore any integrable Hamiltonian sharing the same eigenstates, will lead the system to identical long-time averages. Underlying this assumption of reduction to the diagonal ensemble is the fact that no systematic degeneracies occur so that $E_n \ne E_m$ for any $n \ne m$, i.e.: that we have a simple spectrum. Although technically, for a given system, level crossings can occur so that for certain specific values of $V$ it is possible to find two states sharing a common eigenvalue, no systematic degeneracies occur since sets $\{\Lambda_1 \dots \Lambda_N\}$ solution to the quadratic Bethe equations are all distinct. 

In this work, small enough systems are used so that the sum over the full Hilbert space can be performed completly. However, for larger systems it can also become possible to use Monte Carlo sampling in order to evaluate the involved sums \cite{{montecarlo_calabrese},{montecarlo_gotz},{montecarlo_scardic},{centralfaridet1},{centralfaridet2}}.

The equivalence between long time averages of local operators and the actual properties of the steady-state requires first that the system be large enough so that its eigenspectrum contains a large number of incommensurate frequencies. It also requires a sufficiently spread out projection of the initial state over the eigenstates of the Hilbert space, i.e. the Inverse Participation Ratio: $\sum_n\left|\left< n\right|\left. \psi_{0}\right>\right|^4 \ll 1$ in order for a large number of eigenstates to be populated leading to a large number of contributing frequencies. If these two conditions are met every oscillating term in $\left< \hat{O} \right>(t)$ will then rapidly go out of phase and their sum will cancel out therefore leading to a steady-state whose properties actually correspond to the long time average:

\bea
\lim_{t \rightarrow \infty}\left< \hat{O} \right>(t) = \lim_{t \rightarrow \infty}\frac{1}{t}\int_{0}^{t}\left< \hat{O} \right>(t)dt.
\eea

This evidently excludes systems for which persistent oscillations could remain at long times so that  $\lim_{t \rightarrow \infty}\left< \hat{O} \right>(t)$ is actually no longer properly defined. While for any finite size systems, the true complete dynamics will necesserily retain "weak" fluctuations around the long-time average due to the finite number of frequencies involved, it was shown in \cite{straeter} looking at the bosonic occupation number, that the relaxation towards a well-defined steady-state with well-defined asymptotic expectation values, seems to be a reasonable assumption in this type of models, at least for strongly excited finite sized systems.





\section{Numerical results}
  \label{numresults}

The systems we considered are those composed of a collection of $N=15$ spins-$\frac{1}{2}$ and a boson at frequency $\omega$. The $\epsilon$ parameters are chosen as $\epsilon_i = i$ and therefore, in the corresponding Tavis-Cummings model (eq. \ref{dickeh}), the resulting individual  Zeeman gaps will range from 0 to 14 defining  a band of such gaps with width  $\Delta =14 $. 

At $t=0$, spins will be in a given product state initial configuration (each being either $\uparrow$ or $\downarrow$) and the system has a given number $M$ of total excitations. This number of excitations is conserved over time and corresponds to $M = N_\uparrow +N_b$, i.e. the number of bosons initially in the mode plus the number of spins initially pointing up.

As time evolves, one can expect the interactions to redistribute the excess energy over the various degrees of freedom leading to a loss of z-magnetisation (an in-plane polarising) of each individual spin. However, the integrability limits the freedom of redistribution by imposing additional conserved quantities which raises the question of whether or not an infinite time is sufficient for the various redistribution processes to lead to a steady-state in which the spins have all reached zero-magnetisation along the z-axis. This particular zero-magnetisation state, characterised by a quantum-statistical superposition of both the up and down local spin states, therefore results in a coherent superposition of multiple bosonic occupation states and will therefore be dubbed: superradiant-like steady-state, for the remainder of this paper. The superradiant phase of light-matter quantum systems at large enough couplings is still actively studied: in \cite{transition_nature} the question of whether this transition is a quantum phase transition is answered for Dicke models and Tavis-Cummings models, in \cite{supresstransition} it is shown that the dephasing of the individual atoms destroys the phase transition, in \cite{transversecavity} the phase diagram for ultracold fermions in a transversely pumped cavity is given and in \cite{beyondrwa} the chosen values for the parameters of the system are extreme (the frequency of the two level systems very small compared to the bosonic frequency and the coupling is very strong) to study what happens beyond the RWA.

The following numerical study, will try to shed light on how parameters of the hamiltonian ($\omega$ and $V$) as well as different initial states (spin configurations and number of excitations $M$) affect the resulting steady-states which emerge at long time. A qualitative description of the impact of these various parameters on the steady-states will be proposed and validated through the systematic study of their individual effect.

\subsection{Reference state and coupling constant}
\label{ref_coupling}

This first subsection aims to illustrate that, in the steady-state, there remains a signature of the initial configuration (a common feature of integrable models which has been studied with the Generalised Gibbs Ensemble (GGE) \cite{{RigolGGE},{Heisenbergwin}} and more recently with the Quench Action method \cite{caux_quench_action} which is often use to put the GGE \cite{{Heisenbergfail1},{Heisenbergfail2},{quench_action_harmonics_chains}} to the proof) for any  finite coupling constant. However, at large enough $V$ the steady-state does correspond to the superradiant-like state where each spin has fully plane-polarised. At lower values of $V$, the resulting steady-state can be qualitatively understood as if it resulted from an interrupted relaxation towards the superradiant-like state. 

We present, in the following, the steady-state of local z-magnetisation of each individual spin for a variety of coupling constants starting from an initial spin state chosen to be a typical, i.e. highly entropic, product state. Indeed, since the individual gaps are $\epsilon_i = i$, flipping up the even numbered spins ensures a spreading of the excitations over every available energy scale. One can therefore expect that this highly entropic spin configuration should allow a maximally efficient redistribution of the local excess of energy leading, as easily as possible, to a steady-state which has lost as much information as possible about the initial state.

\begin{figure}[h]\includegraphics[width=\textwidth]{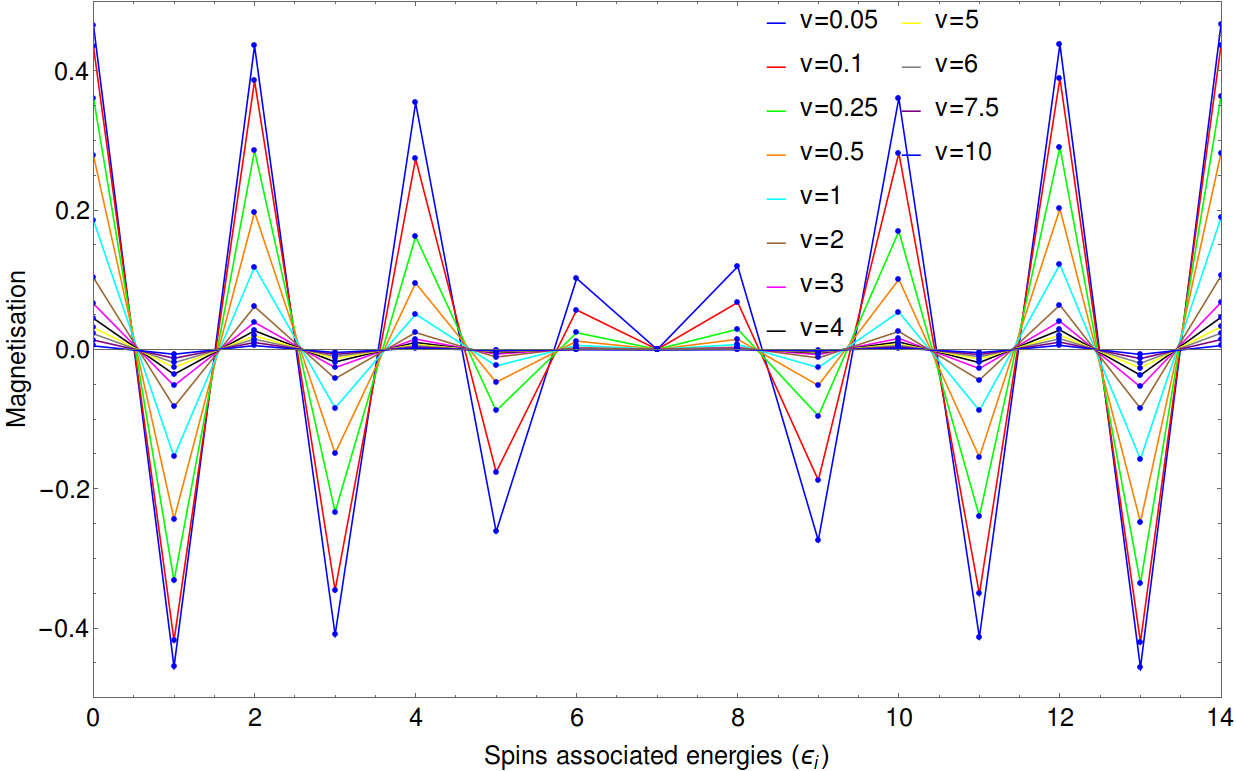}  
\vspace{-0.5cm}
\caption{Local magnetisation in the steady-state for the initial state: $\left|17;\uparrow\downarrow\uparrow\downarrow\uparrow\downarrow\uparrow\downarrow\uparrow\downarrow\uparrow\downarrow\uparrow\downarrow\uparrow \right>$ for a variety of coupling constants $V$. The total number of excitations is $M=25$ and the bosonic frequency is set to $\omega=6.95$, very close to the middle of the energy band $\Delta =14$.} 
\label{reference}
\end{figure}

As is obvious from these numerical results, the local magnetisations which are found in the steady-state are strongly influenced by the value of the coupling constants $V$. Even given an infinite time to do so, the exchange processes between the spins and the bosonic mode, which can occur when the system dynamics is driven to its steady-state by $H_{TC}$ for example, are unable to establish the superradiant-like steady-state one might have naively expected. 

Two distinct characteristic energy scales can be defined on the spin side of the model, namely the full bandwidth $\Delta =14$ of the inhomogeneous Zeeman gaps as well as the separation between these individual gaps $\delta = \epsilon_{i+1} - \epsilon_i = 1$, fixing the largest spin-spin coupling which appears in the conserved charges (\ref{consboson}) of this class of models. Introducing the dimensionless parameter $\frac{\Omega}{\Delta} \equiv \frac{V \sqrt{N}}{\Delta}$, the ratio between the Rabi frequency and the bandwidth, one sees that only in a strong coupling limit $\frac{\Omega}{\Delta} > 1$ we do find a steady-state which has z-magnetisation $\approx 0$ for every spin. Below these coupling strengths ($V < \ \ \sim 3.61$), a strong signature of the initial state is maintained in the steady-state since individual spins retain an average magnetisation in the direction in which they were initially polarised.

To simplify the discussion, we numerically define a value $V_{sup}$ as the coupling for which the steady-state average local magnetisation is $\left<S^z_i\right> <10^{-2} \ \forall \ i$. For the situation presented in Fig.\ref{reference}, this condition is met for the blue curve ($V=10$), which sets $V_{sup}\approx 10$ for this particular initial configuration. At $V \ge V_{sup}$ the system will systematically reach such a common superradiant-like steady-state which has lost any trace of the initial spin configuration. However, for $V<V_{sup}$ these first results seems to indicate that one can think, qualitatively, in a pseudo-dynamical mindset where $V$ plays the role of time in a relaxation process from the initial configuration towards the zero-magnetisation superradiant-like state. Indeed, as $V$ grows, the steady-state local magnetisations are systematically moving away from their initial values, going towards the value 0, which will only be reached at a strong enough $V \ge V_{sup}$.

Within this pseudo-dynamical description, one can now try to define how a variety of factors will influence $V_{sup}$. The following subsections will therefore explore the impact of the value of the bosonic frequency $\omega$, of various initial spin configurations and of the initial number of bosonic excitations on the steady-states reached for a given $V$.



Since this reference initial state presented in Fig.\ref{reference} is highly entropic in the spin sector, we can expect $V_{sup}$ to be smaller than for other less entropic initial spin configurations. Thinking of the dynamics as a sequence of individual exchange processes which ultimately lead to the redistribution of the initially local excitations, the energetic proximity of a large number of states (high entropy) should make this redistribution more efficient.  We can also presuppose that another aspect which could enhance this efficiency is to have an initial state which has a lot of excitations available, i.e. a large number of bosons at $t=0$. Therefore, increasing $M$ could help the system to reach a superradiant-like steady-state, in other words, to lower $V_{sup}$ (see subsection \ref{excitationM}) because it has at its disposal more bosons to help the redistribution.

One last aspect visible in Fig.\ref{reference} is the apparent impact on a given spin of having its associated energy $\epsilon_i$ near the frequency of the bosonic mode $\omega$. Indeed, we can clearly notice in Fig.\ref{reference} that spins with $\epsilon_i$ close to $\omega$ reach a $0$ z-magnetisation steady-state for much weaker coupling than those far from $\omega$. For $\epsilon_7 \ (\approx \omega)$, we can even assume that any non-zero coupling is enough for it to be polarised in the plane at $t\rightarrow \infty$. This appears to be consistent with a resonance phenomenon which for resonant energies lead to highly efficient exchange processes. This can be easily understood for a system evolving under a TC Hamiltonian (eq. \ref{dickeh}) in which exchange processes only flip various spins through the creation or absorption of a bosonic excitation. However, we remind the readers, that the results presented here are valid for an evolution driven by any Hamiltonian of the integrable Richardson-Gaudin class considered here. Choosing a specific $R_i$ (\ref{consboson}) as the Hamiltonian responsible for the time-evolution makes this resonance effect much more subtle to understand since a single spin is then directly coupled to the bosonic mode, while other spins only have exchange processes which couple them directly to one specific spin: $S_i$.




\begin{figure}[h]\includegraphics[width=\textwidth]{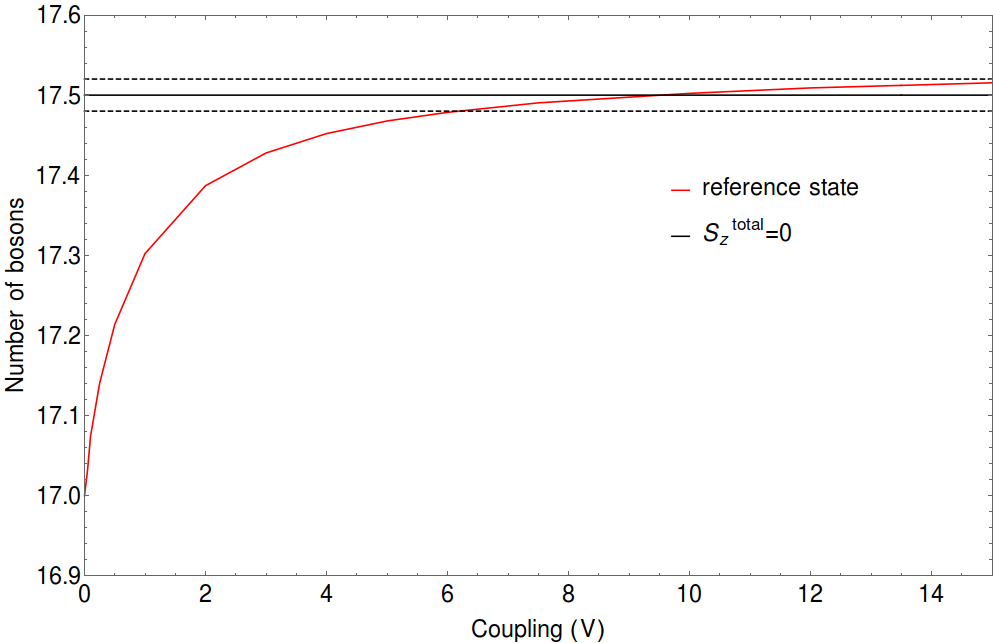} 
\caption{Bosonic occupation in the steady-state for the initial state: $\left|17;\uparrow\downarrow\uparrow\downarrow\uparrow\downarrow\uparrow\downarrow\uparrow\downarrow\uparrow\downarrow\uparrow\downarrow\uparrow \right>$ as a function of $V$. The total number of excitations is $M=25$ and the bosonic frequency is set to $\omega=6.95$ in the middle of the energy band.} 
\label{bosonreference}
\end{figure}

Before moving to the next section which will validate the assumptions made here, we first present the bosonic occupation in the state for the case at hand. The total number of excitations $M=\left<b^{\dag}b\right>+\left<S_{z}^{tot}\right>$ being constant, this quantity is completely related to the total magnetisation and can serve as a quantity describing in global terms the proximity to a $0$ average z-magnetisation steady-state. In Fig.\ref{bosonreference}, we observe that the average number of bosons almost stops changing so that the various stead-states mostly differ by a reorganisation in the spin sector. For this reference initial state, the steady-state bosonic occupation reaches $\left<b^{\dag}b\right> = 17.5 \pm 0.02$ (dashed lines) for $V\approx 6$. In this scenario, considering the conservation of the total number of excitations, $\left<b^{\dag}b\right>=17.5$ (black line) would correspond to a state where the total magnetisation along the z axis cancels, i.e. the spins are all half-excited. One can also clearly notice a weak overshoot ($\left<b^{\dag}b\right> > 17.5$) for $V=10$ which contradicts the hypothesis of a monotonic evolution (in $V$) towards a fully plane polarised steady-state. We will therefore verify in the upcoming sections that this overshoot will be dampened when the coupling is pushed to higher values, leading us back towards the zero-average magnetisation steady-state.


 \subsection{Entropy in the spin sector}
 
In order to demonstrate the claim that a highly entropic initial state favours a superradiant-like steady-state, we now look at a radically different initial spin configuration explicitly chosen to have low entropy in the spin sector. Having every spin down polarised and pushing the bosonic frequency near the edge of the available spin gaps, we create a configuration which, at low coupling, would be at very high energy in the global TC Hamiltonian spectrum. Lying close to the edge of the spectrum, the low density of states should prevent efficient relaxation.

\begin{figure}[h]\includegraphics[scale=0.35]{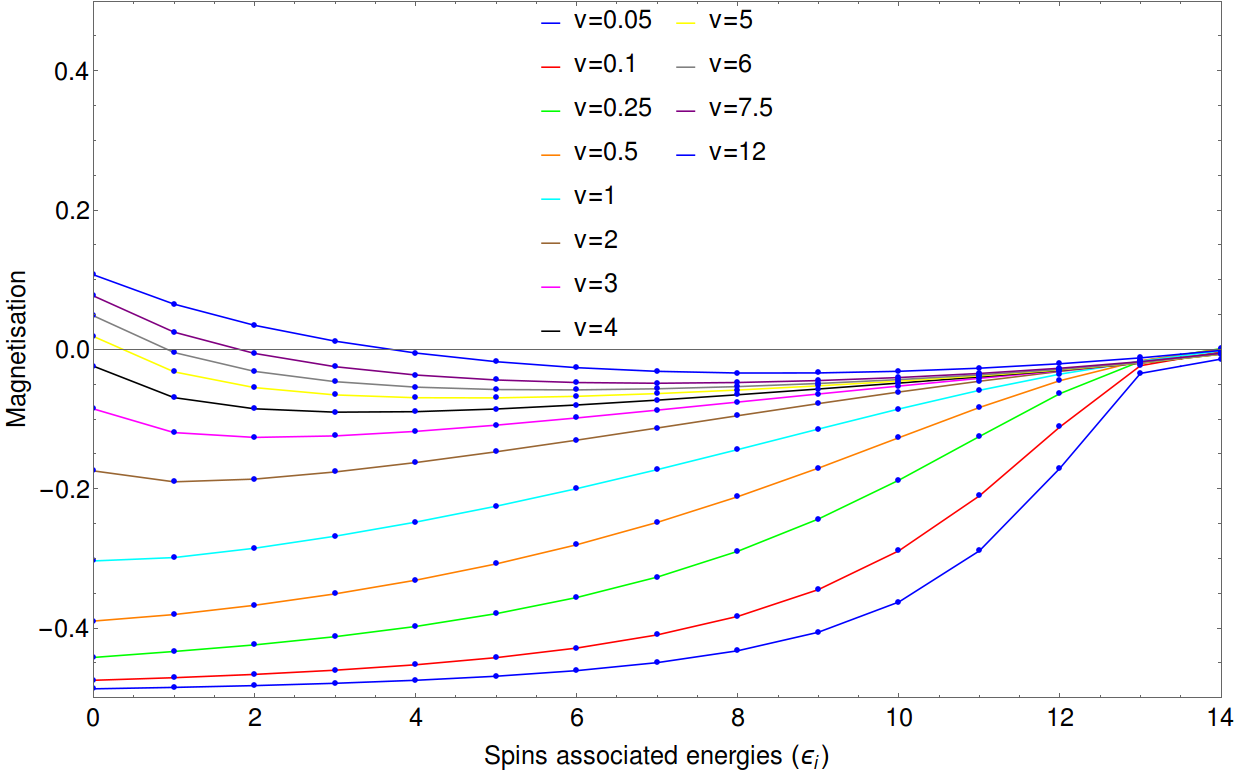} 
\vspace{0.1cm}
\caption{Local magnetisation in the steady-state for the initial state: $\left|25;\downarrow\downarrow\downarrow\downarrow\downarrow\downarrow\downarrow\downarrow\downarrow\downarrow\downarrow\downarrow\downarrow\downarrow\downarrow \right>$ as a function of $V$. The total number of excitations is $M=25$ and the bosonic frequency is set to $\omega=13.5$ near the maximum of the energy band.} 
\label{Szi1525150}
\end{figure}

Indeed, as seen in Fig.\ref{Szi1525150}, even with a coupling of $V=12$, which is above the reference state's $V_{sup}$, spins have not yet plane-polarised, confirming the assumption that highly entropic  initial spin configurations lead to smaller $V_{sup}$.  There remains to be seen whether stronger coupling constants will indeed lead to a zero average magnetisation steady-state, especially considering the strong overshoot  $\left< S^z_i\right> > 0$ seen for the low-index spins far from resonance ($\epsilon_i$ far from $\omega$) seen at couplings  $V> 5$. 


\begin{figure}[h]\includegraphics[scale=0.35]{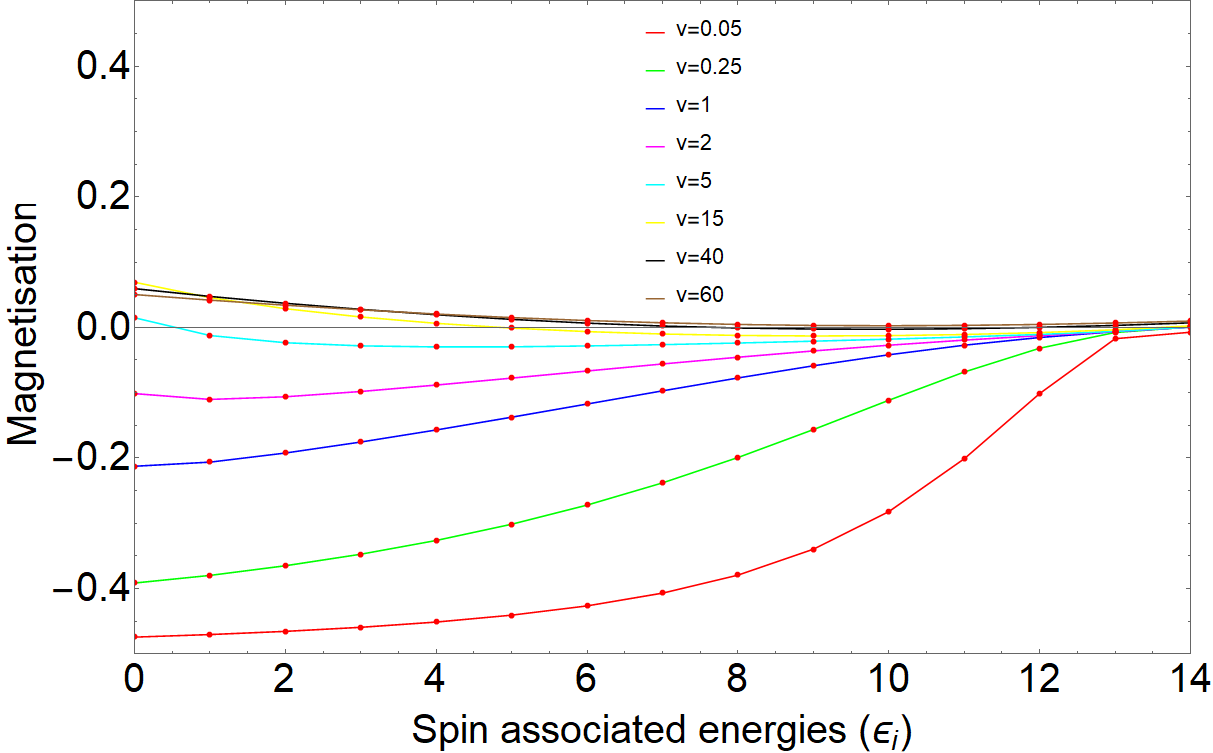} 
\vspace{0.1cm}
\caption{Local magnetisation in the steady-state for the initial state: $\left|50;\downarrow\downarrow\downarrow\downarrow\downarrow\downarrow\downarrow\downarrow\downarrow\downarrow\downarrow\downarrow\downarrow\downarrow\downarrow \right>$ as a function of $V$. The total number of excitations is $M=50$ and the bosonic frequency is set to $\omega=13.5$ near the maximum of the energy band.} 
\label{Szi1550150}
\end{figure}

Fig.\ref{Szi1550150} completes the previous data by showing $\left<S^z_i\right>$ for a wider range of couplings now ranging from $V=0.05$ to $V=60$. The total number of excitations has also been pushed higher to $M=50$ which, as mentioned in the previous section (and explicitly studied in section \ref{excitationM}), should favour a zero-magnetisation steady-state at lower values of $V$.

By first comparing, in Fig.\ref{Szi1550150} and in Fig.\ref{Szi1525150}, the local steady-state magnetisation of spin $i=0$ at $V=1$, for example, we do see a first clear confirmation that a larger number of excitations $M$ leads, for a given spin, to a steady-state with average magnetisation closer to $0$ at a given value of $V$. On the other hand, we do confirm in this data set that the overshoot which leads to $\left<S^z_i\right> > 0$ for the spins far from resonance is indeed reduced when going to much stronger couplings. It is consistent with the hypothesis that, at large enough coupling $V$ the steady-state will systematically have lost initial state information, therefore reaching the zero-magnetisation configuration for arbitrary initial states. However, comparison with the previous section's highly entropic reference state clearly demonstrates that such a low entropy initial state will require much larger values of couplings for this universal steady-state to emerge.

From the data presented in those sections, the qualitative description of the local magnetisation properties remains consistent with the interpretation that the system "relaxes" towards this universal superradiant-like steady-state, with the coupling $V$ playing the role of time in this pseudo-dynamical description. The variety of factors which control the efficiency of this relaxation, i.e. how large $V$ has to be in order for this universal steady-state to be realised, will be systematically studied in the next few sections.

\subsection{Resonance with the bosonic mode $\omega$}

As described in the previous sections, it appears that spins which are nearly resonant with the bosonic frequency  ($\epsilon_i \approx \omega$) reach zero magnetisation in the steady-state for much weaker couplings than those far from resonance. In order to explicitly confirm this fact, the next figure shows the local magnetisation in the steady-state in four different cases which only differ by their value of bosonic frequency, $\omega=13.5,10.5,6.95$ and $0.5$. In each case, the initial states are identical, with spins initially fully down polarised states while the bosonic modes contains 25 excitations. 

\begin{figure}[h]
   
    \centering
 \begin{tabular}{cc}
 \includegraphics[width=0.55\linewidth]{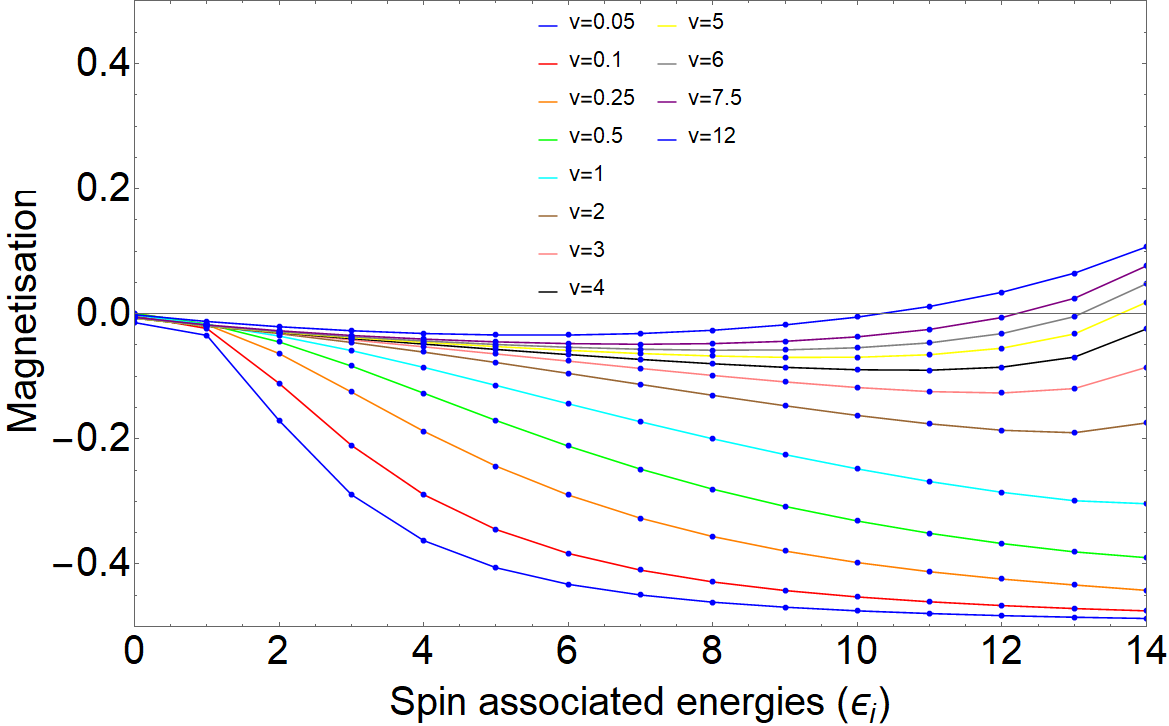}    & \includegraphics[width=0.55\linewidth]{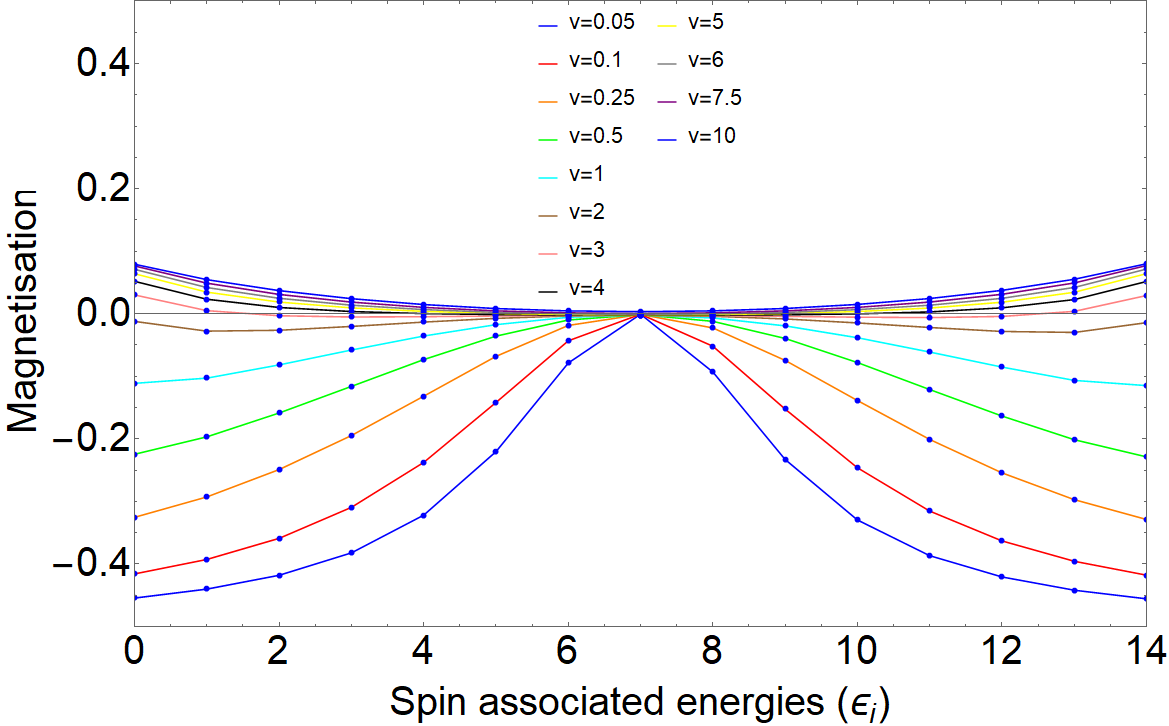} \\
 a) $\omega=0.5$ & b) $\omega = 6.95$ \\[1em]
 \includegraphics[width=0.55\linewidth]{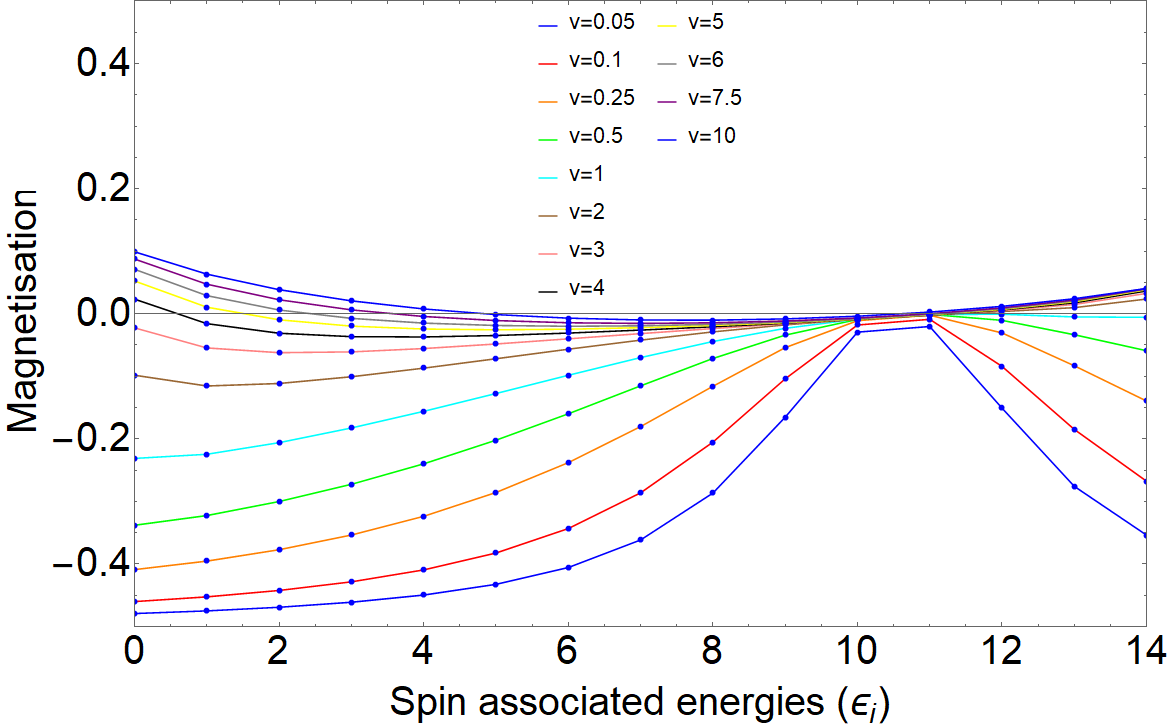}    & \includegraphics[width=0.55\linewidth]{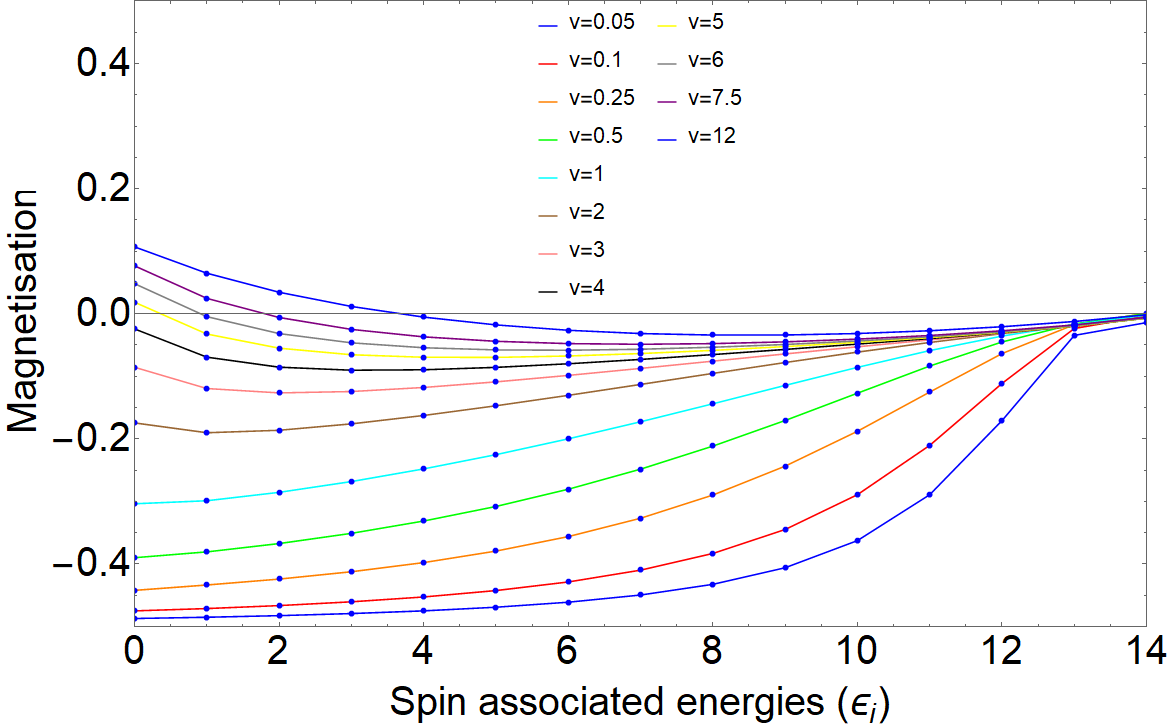} \\
 c) $\omega = 10.5$ & d) $\omega = 13.5$ \\[1em]
 \end{tabular}
 \caption[Local magnetisation in steady-states for the initial state: $\left|25;\downarrow\downarrow\downarrow\downarrow\downarrow\downarrow\downarrow\downarrow\downarrow\downarrow\downarrow\downarrow\downarrow\downarrow\downarrow \right>$ for a variety of coupling constants $V$ and four different bosonic frequencies. The total number of excitations is fixed at $M=25$.]{Local magnetisation in steady-states for the initial state: $\left|25;\downarrow\downarrow\downarrow\downarrow\downarrow\downarrow\downarrow\downarrow\downarrow\downarrow\downarrow\downarrow\downarrow\downarrow\downarrow \right>$ for a variety of coupling constants $V$ and four different bosonic frequencies. The total number of excitations is fixed at $M=25$.}
 \label{fig:reso1}
 \end{figure}

First of all, in the four groups of plots presented (Fig.\ref{fig:reso1}) the local magnetisation never reaches the global criterion $\left<S^z_i\right> <10^{-2}\forall i$ defining $V_{sup}$, which is not surprising since the initial spin configuration chosen here has low entropy and should therefore require $V >> 10$ to reach the zero-magnetisation steady-state. Panel b) can also be compared to the reference state of Fig.\ref{reference} since both share the same $\omega$ and differ only in the initial configuration. This comparison confirms, looking at purely initial state entropic-effect, that $V_{sup}$ is now $>> 12$ and therefore does become larger in this low entropy case.

Comparing the four panels in Fig.\ref{fig:reso1}, it becomes remarkably clear that the coupling is much more effective in plane-polarising spins whose $\epsilon_i \approx \omega$. The effect is particularly clear on panel b) in which spin 7 has $ \epsilon_7 - \omega = 0.05$, but is also clearly evident in all other panels in which the two specific spins closer to resonance have $\epsilon_k - \omega = 0.5$.

While, in a purely TC-like model, this type of resonant behaviour makes perfect sense since spin-flips whose gap corresponds to the bosonic frequency would be favoured, one needs to keep in mind the fact these results also apply for a "central spin" Hamiltonian $R_i$ where only one central spin is directly coupled to the bosonic mode. Indeed, in the TC case the bath of bosons can continuously exchange excitations directly with each individual spin and these processes become very effective at the resonance. In a central spin case, the bosonic bath can only flip one specific spin and would then need to wait for the central spin to flip back down due to its interaction with the remaining spins before another bosonic excitation can be used again to flip the central spin up. The real time dynamics in both cases would therefore be radically different but the same steady-state would be reached, steady-state which shows the same pseudo-dynamical resonant effect: spins with $\epsilon_i \approx \omega$ have a zero average z-magnetisation in the steady-state even at very weak coupling, while spins far from resonance require higher couplings for their steady-state to show the same property.


While one could have expected the local magnetisation to exhibit a monotonic behaviour in $\omega - \epsilon_i$, it is particularly clear for $V=2$ in panel (b) that this is not the case, since the extreme spins ($i=0$ and $i=14$) show a steady-state with an average magnetisation closer to zero than the one of the next spins ($i=1$ and $i=13$). This non-monotonic behaviour in $\omega - \epsilon_i$ does not find a simple intuitive explanation in the TC model but, when considering the other underlying conserved charges $R_i$, the presence of the spin-spin coupling terms $\frac {\vec{S}_i \cdot \vec{S}_j}{\epsilon_i-\epsilon_i} $ should make it clear that the distance from the resonance $\omega - \epsilon_i$ is not the only parameter controlling individual spins. Indeed, the set of conserved quantities, which controls the redistribution of local excitations, does show an explicit dependence on an individual spin's $\epsilon_i$ "position" compared to the other spins' $\epsilon_j$ ultimately affecting its individual relaxation behaviour.

Any given integrable model in the class studied here will ultimately show steady-states properties which are not intuitively seen in the Hamiltonian itself. Be it this non-monotonic behaviour in $\omega - \epsilon_i$ not easily inferred from $H_{TC}$, or conversely the resonance effect ($\epsilon_i \approx \omega$) not easily seen in one of the central spin Hamiltonians $R_j$. While their real-time dynamics, i.e. the way they reach the steady-state, would differ for distinct Hamiltonians, the steady-states themselves would retain the same properties for each and everyone.


\begin{figure}[h]\hspace{1cm}\includegraphics[scale=0.35]{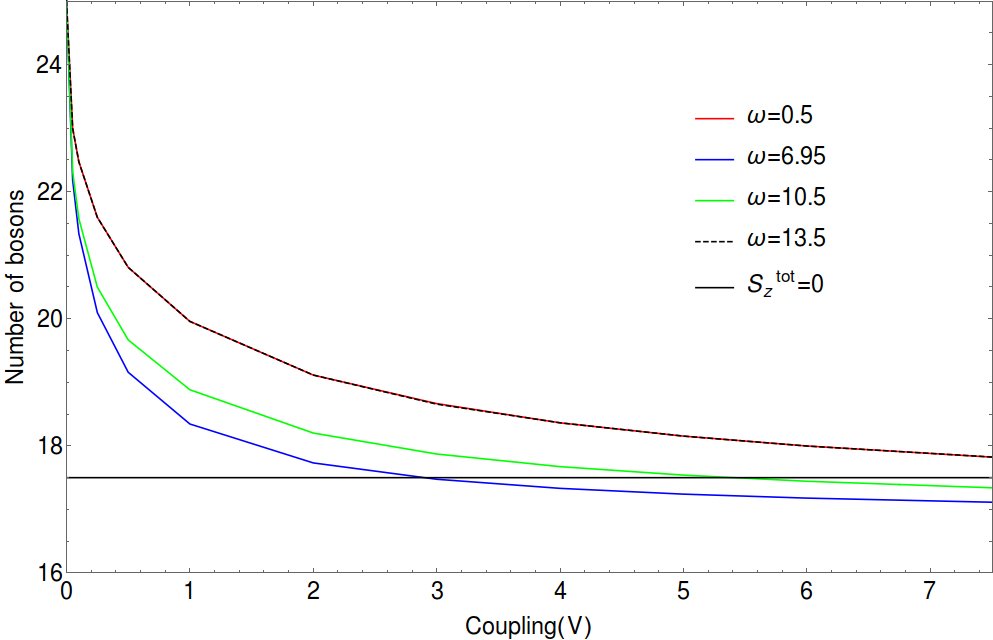}  
\vspace{-0.0cm}
\caption{Bosonic occupation in steady-states for the initial state: $\left|25;\downarrow\downarrow\downarrow\downarrow\downarrow\downarrow\downarrow\downarrow\downarrow\downarrow\downarrow\downarrow\downarrow\downarrow\downarrow \right>$ for a variety of coupling constants $V$. The total number of excitations is $M=25$ and the bosonic frequency are set to $\omega=0.5,6.95,10.5,13.5$.} 
\label{bosonresonance}
\end{figure}

In Fig.\ref{bosonresonance} the steady-state bosonic occupation is plotted for the four same situations which only differ by the values of $\omega$. The curves for $\omega=0.5$ and $13.5$ (red and dashed black ones) in Fig.\ref{bosonresonance} are identical due to the symmetry between these two cases, for which the roles of the spin of index $i$ and of the one of index $N-i$ are exchanged. The two other curves (blue and green) clearly cross the black line which corresponds to the bosonic occupation found for zero total spin magnetisation. This undershoot in bosons is complementary to the overshoots in magnetisation seen in Fig.\ref{fig:reso1}. In the previous section, we saw that such an overshoot in magnetisation can occur for lower $V$ than the one needed for the spins to become plane-polarised at larger couplings. One can therefore assume that the lower the $V$ for which this undershoot occurs, the lower $V_{sup}$ should be as well. Since the ($\omega = 6.95$) is the first to do so, this becomes consistent with the conclusion that the further $\omega$ is from the middle of the energy band, the greater $V_{sup}$ should be. The idea is also coherent with the resonance in $\omega-\epsilon_i$ idea, in that a bosonic frequency placed in the middle of the band will lower the largest possible $\omega-\epsilon_i$ therefore favouring in-plane polarisation at lower couplings. As is clearly seen in Fig.\ref{fig:reso1}, it is indeed the spins farther off-resonance which, at larger couplings, show the strongest deviations for the zero-average z-magnetisation which defines the universal large coupling steady-state.


\begin{figure}[h]\hspace{1cm}\includegraphics[scale=0.3]{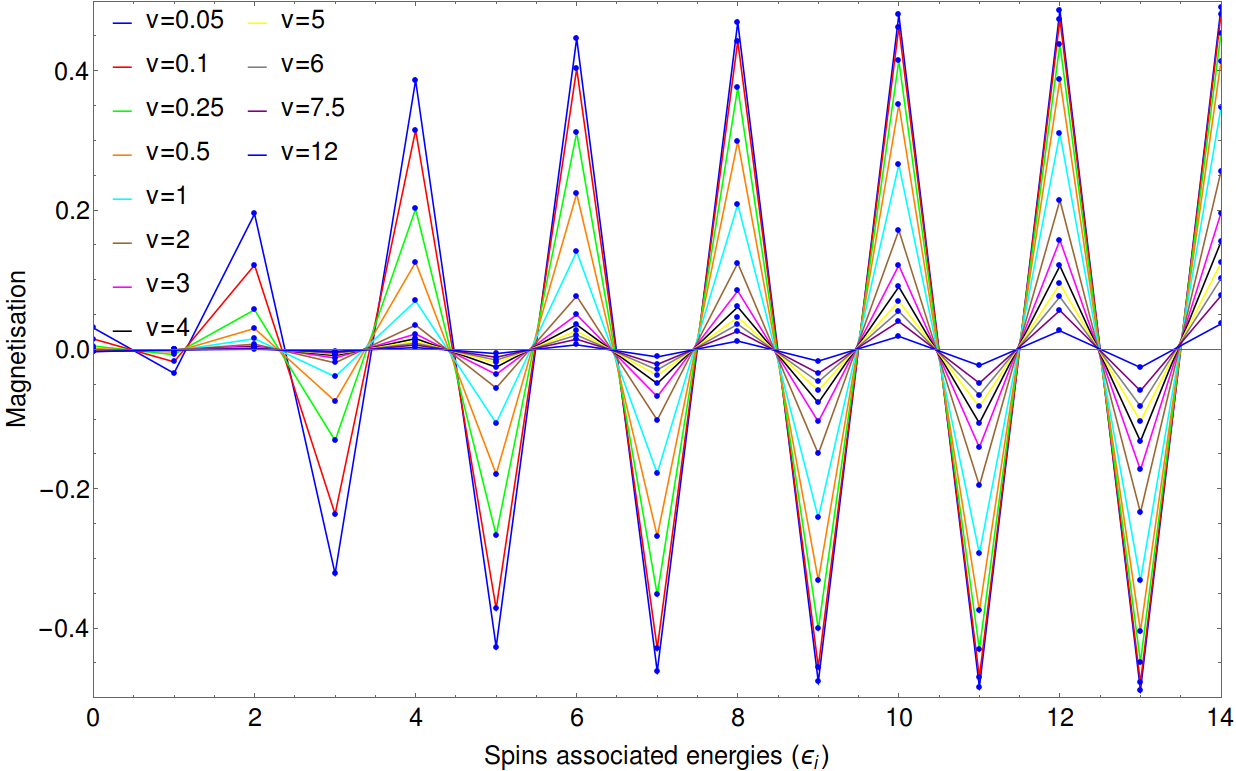}  
\vspace{-0.0cm}
\caption{Local magnetisation in the steady-state for the initial state: $\left|17;\uparrow\downarrow\uparrow\downarrow\uparrow\downarrow\uparrow\downarrow\uparrow\downarrow\uparrow\downarrow\uparrow\downarrow\uparrow \right>$ for a variety of coupling constants $V$. The total number of excitations is $M=25$ and the bosonic frequency is set to $\omega=0.5$, near an edge of the energy band.} 
\label{szi1525entmax2}
\end{figure}

The plots in Fig.\ref{szi1525entmax2}, showing the local magnetisation for the same initial spin configuration as in the reference case, further confirms the picture defined in this subsection. Only $\omega$ (here it is equal to $0.5$) differs from the reference plot Fig.\ref{reference} and comparing both, it is clear that the system has not reached the fully plane polarised state for $V=12$. Despite the alternating spin configuration, adding initial entropy to the system, the off-resonance spins (at $\epsilon_{i}$ far from $\omega$) still show a much stronger memory of their initial orientation at a coupling strength for which this memory had disappeared in Fig.\ref{reference}.

\subsection{Initial spin configuration}
\label{initstate}

\begin{figure}[h]
   
    \centering
 \begin{tabular}{cc}
 \includegraphics[width=0.55\linewidth]{2moyenszinf1525150.png}    & \includegraphics[width=0.55\linewidth]{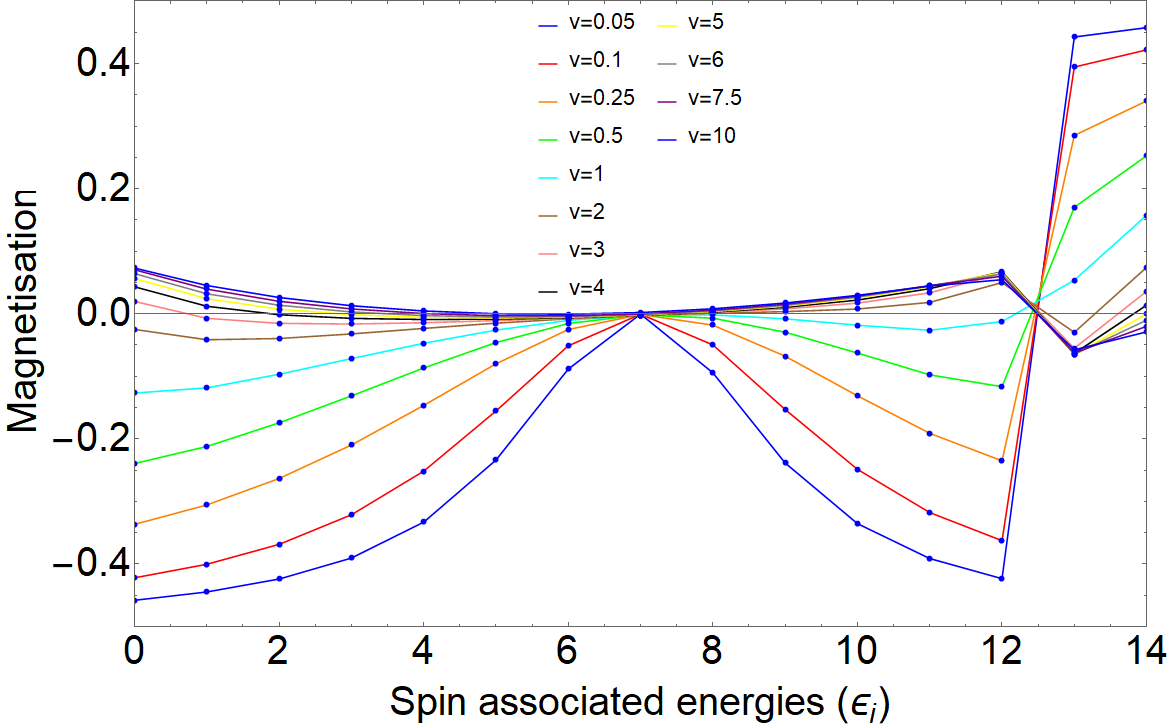} \\
 a) Initial state: $\left|25;\downarrow\downarrow\downarrow\downarrow\downarrow\downarrow\downarrow\downarrow\downarrow\downarrow\downarrow\downarrow\downarrow\downarrow\downarrow \right>$ & b) Initial state: $\left|23;\downarrow\downarrow\downarrow\downarrow\downarrow\downarrow\downarrow\downarrow\downarrow\downarrow\downarrow\downarrow\downarrow\uparrow\uparrow \right>$ \\[1em]
 \includegraphics[width=0.55\linewidth]{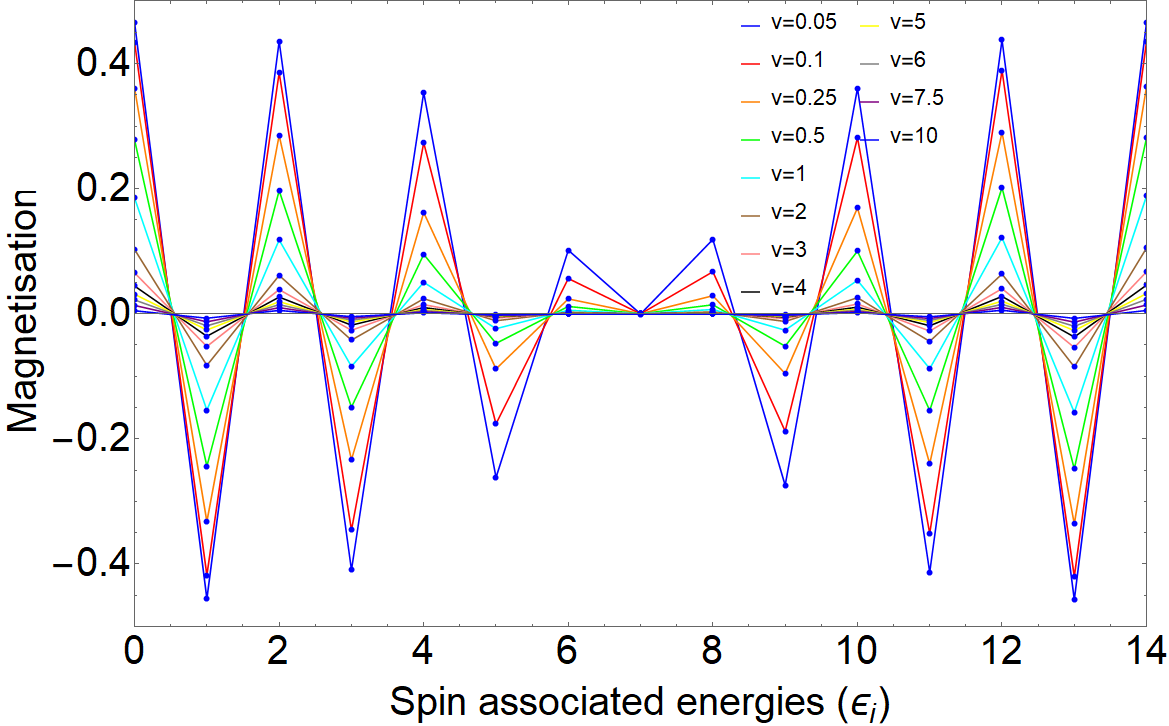}    & \includegraphics[width=0.55\linewidth]{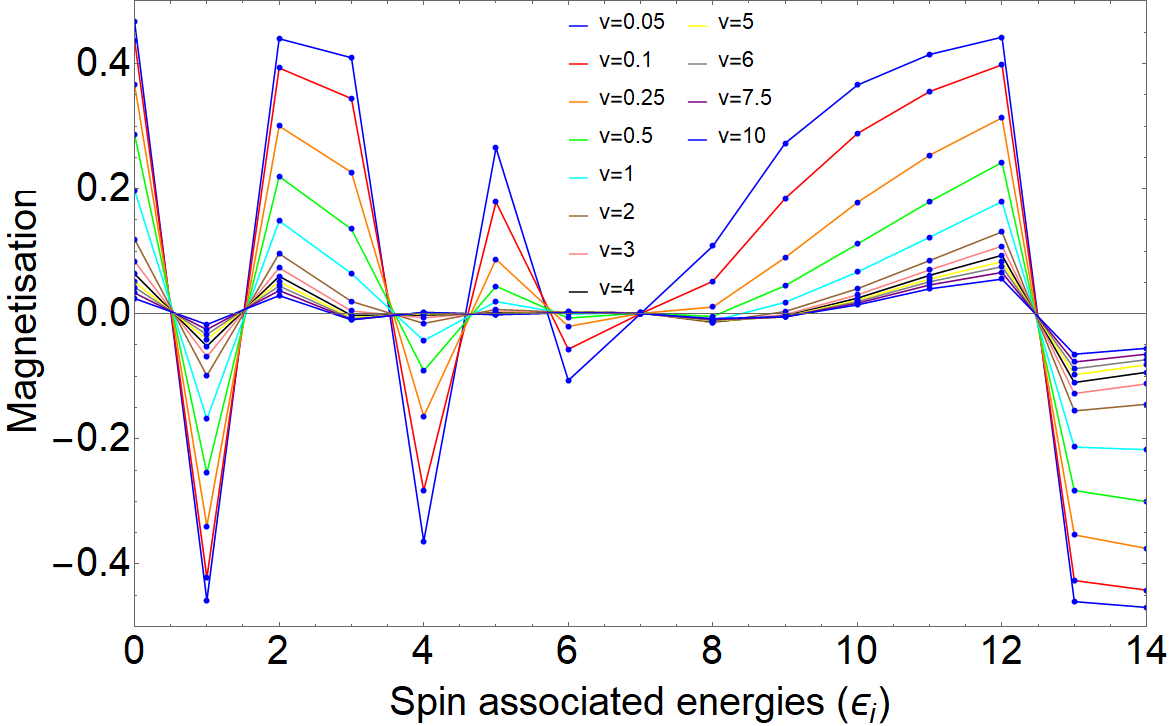} \\
 c) Initial state: $\left|17;\uparrow\downarrow\uparrow\downarrow\uparrow\downarrow\uparrow\downarrow\uparrow\downarrow\uparrow\downarrow\uparrow\downarrow\uparrow \right>$ & d) Initial state: $\left|16;\uparrow\downarrow\uparrow\uparrow\downarrow\uparrow\downarrow\downarrow\uparrow\uparrow\uparrow\uparrow\uparrow\downarrow\downarrow \right>$ \\[1em]
 \end{tabular}
 \caption[Local magnetisation in steady-states for the initial state: $\left|25;\downarrow\downarrow\downarrow\downarrow\downarrow\downarrow\downarrow\downarrow\downarrow\downarrow\downarrow\downarrow\downarrow\downarrow\downarrow \right>$ for a variety of coupling constants $V$ and four different bosonic frequencies. The total number of excitations is fixed at $M=25$.]{Local magnetisation in steady-states for the initial state: $\left|25;\downarrow\downarrow\downarrow\downarrow\downarrow\downarrow\downarrow\downarrow\downarrow\downarrow\downarrow\downarrow\downarrow\downarrow\downarrow \right>$ for a variety of coupling constants $V$ and four different bosonic frequencies. The total number of excitations is fixed at $M=25$.}
 \label{fig_initialstate}
 \end{figure}

To complete the previous analysis, the results presented in this section look at the impact of "local entropy" in the initial spin configuration. The four graphics of Fig.\ref{fig_initialstate} therefore show the local magnetisation in the steady-state for a variety of initial spin configurations.


In Fig.\ref{fig_initialstate}, panel (c) reproduces the reference result initially presented in Fig.\ref{reference}, panel (a) the low entropic fully polarised initial spin state shown in Fig.\ref{fig:reso1}, while (b) and (d) respectively introduce a single domain wall and larger a number of them in the initial spin configuration. The bosonic frequency is kept fixed in the middle of the energy band, the total number of excitations is also kept fixed at $M=25$. All three states (a),(b) and (d) can be seen as having lower entropy than the alternating spin configuration of the reference panel (c) and, as a result, in both cases $V_{sup}$ is indeed higher than in the reference state ($V_{sup}=10$) since at $V=10$ none of them has a steady-state such that $\left<S^z_i\right> <10^{-2} \ \forall \ i$.

The main differences between the reference initial state (c) and the one in (a) is evidently the total spin magnetisation of the initial state, which are respectively $0.5$ and $-7.5$, and the neighbouring environment of each spin. In the reference state, each neighbouring spin is alternated whereas all the spins are pointing in the same direction in the other case. The total magnetisation of the fully plane polarised state is zero, so one argument to explain why $V_{sup}$ is smaller in the reference (a) than in all the cases presented in this subsection is that the initial state of reference is simply closer in total magnetisation to this zero-magnetisation steady-state. 

 One can now turn to trying to understand the effect of a local alternation of spins, i.e. local entropy. In panel (b), looking at the spin of index $1$ and $13$, for weak to intermediate couplings, it appears that spin 13 needs weaker coupling to erase the initial state memory, i.e. its magnetisation gets closer to zero that the one of spin number 1 despite the fact that they are equally far in energy to the bosonic frequency. It therefore seems reasonable to suppose that the local entropy, i.e. the alternation of spins close to spin 13 is responsible for making $V$ more efficient locally in creating the zero-magnetisation steady-state. In panel (d), comparing the ensemble of spins with indexes $3$, $4$ and $5$ to the one of spins with indexes $9$, $10$ and $11$ confirms this idea that an initial alternated spin configuration favours the plane-polarised steady-state. 
 
 Therefore we understand that, even locally, the configuration of spins in the initial state can have an impact on the memory of the initial state. An initial state which deviates from the strict alternation of spins we called our reference, which is the most favourable initial state, will involve a larger $V_{sup}$ for the memory of the initial configuration to be gone in the steady-state.

\begin{figure}[h]\hspace{1cm}\includegraphics[scale=0.4]{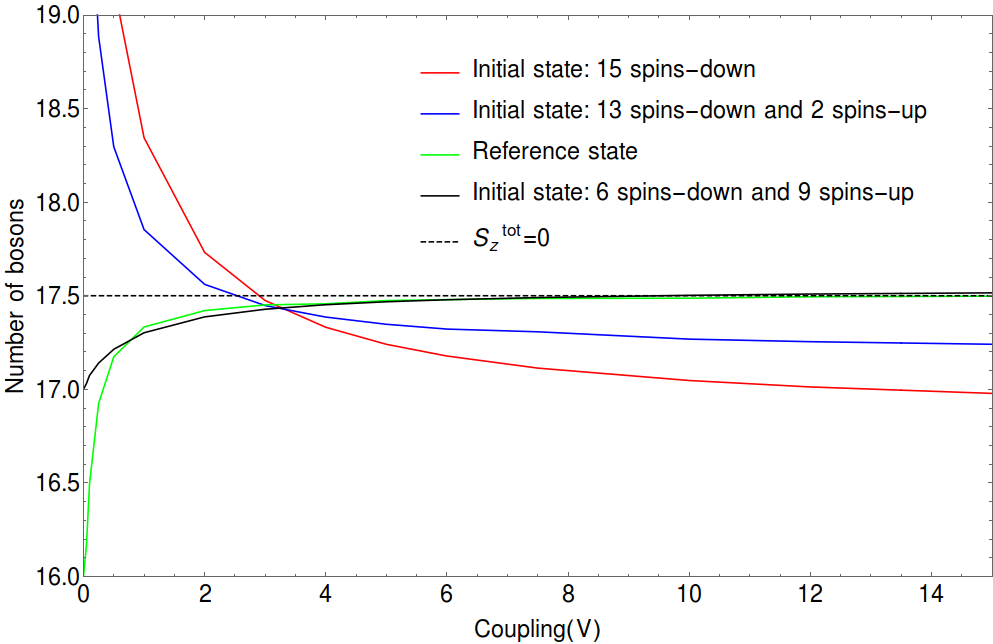}  
\vspace{-0.5cm}
\caption{Bosonic occupation in the steady-state for various initial states: $\left|17;\uparrow\downarrow\uparrow\downarrow\uparrow\downarrow\uparrow\downarrow\uparrow\downarrow\uparrow\downarrow\uparrow\downarrow\uparrow \right>$, $\left|25;\downarrow\downarrow\downarrow\downarrow\downarrow\downarrow\downarrow\downarrow\downarrow\downarrow\downarrow\downarrow\downarrow\downarrow\downarrow \right>$, $\left|23;\downarrow\downarrow\downarrow\downarrow\downarrow\downarrow\downarrow\downarrow\downarrow\downarrow\downarrow\downarrow\downarrow\uparrow\uparrow \right>$ and $\left|16;\uparrow\downarrow\uparrow\uparrow\downarrow\uparrow\downarrow\downarrow\uparrow\uparrow\uparrow\uparrow\uparrow\downarrow\downarrow \right>$ as a function of $V$. The total number of excitations is $M=25$ and the bosonic frequency is set to $\omega=6.95$.}
\label{various_init_boson} 
\end{figure}

 Fig.\ref{various_init_boson}, showing the bosonic occupation in terms of $V$, illustrates the effect of the alternation of spins on the global system. The curves representing the evolution of the bosonic occupation of states containing entropic spin configuration in their initial states (green and black ones) reach their final values, corresponding to a cancelling total magnetisation among the z axis, for $V\approx 8$ whereas the two other curves are still decreasing for $V=15$. This behaviour confirm that an entropic initial state in the spin sector is favouring a fully plane polarised steady-state. Another aspect Fig.\ref{various_init_boson} is that having more bosonic excitations can help the redistribution of excitations at weak coupling since, looking at the curves corresponding with initial state containing more bosonic excitations (blue and red ones), their weak coupling slope is more faster than the ones of the green and black curves. Nonetheless, in that scenario, having more initial bosonic excitations implies having an initial spin configuration with few spin alternations which ultimately  prevents the system from reaching the zero-magnetisation steady-state.


\subsection{Spin flip at the resonance}

\begin{figure}[h]
   
    \centering
 \begin{tabular}{cc}
 \includegraphics[width=0.55\linewidth]{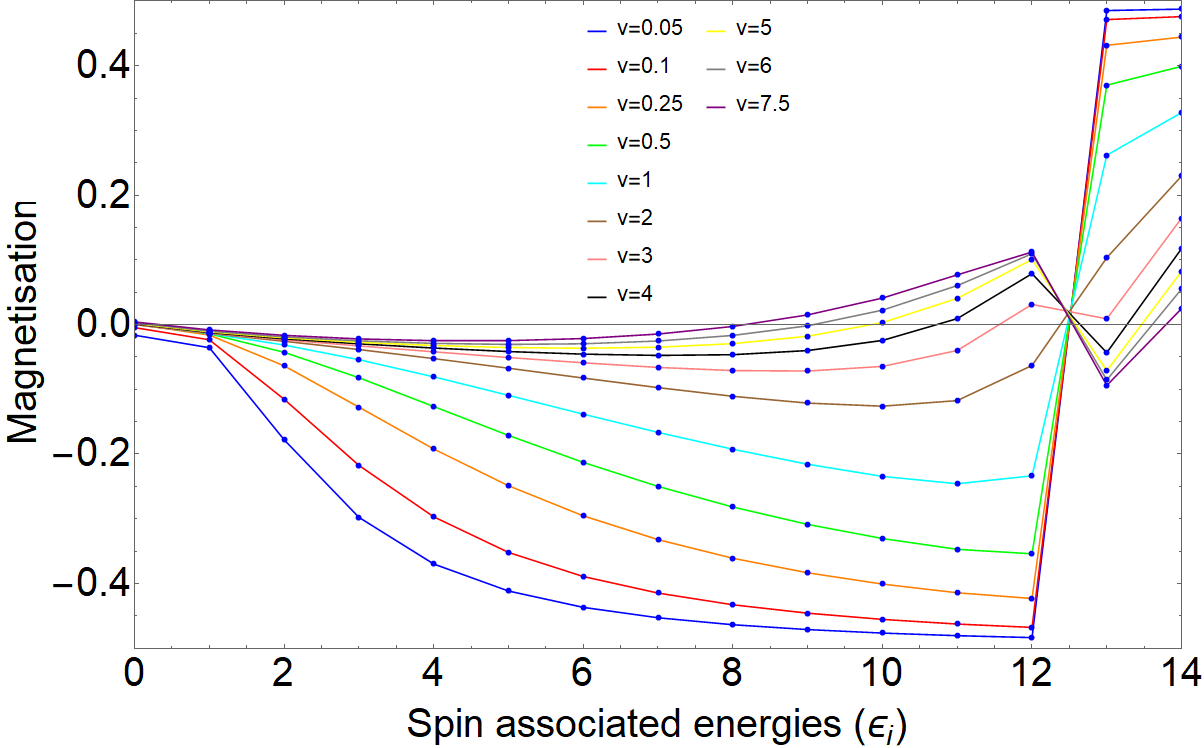}    & \includegraphics[width=0.55\linewidth]{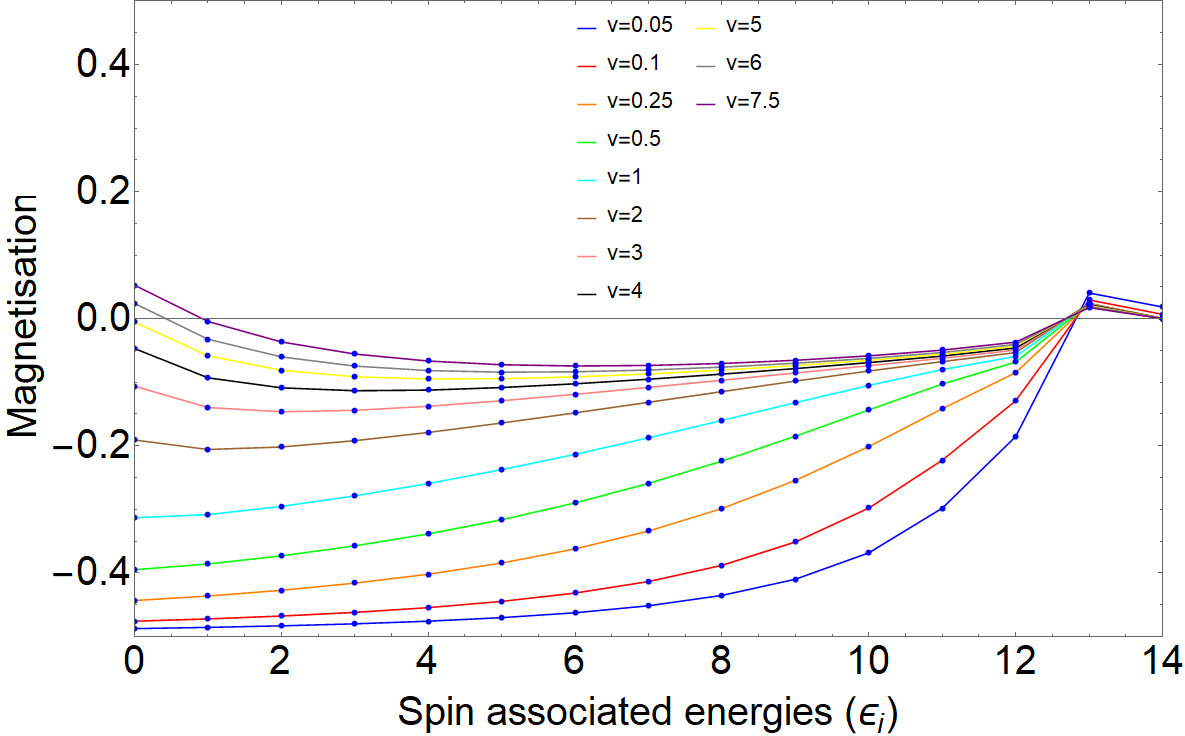} \\
 a) $\omega=0.5$ & b) $\omega=13.5$ \\[1em]
 \end{tabular}
 \caption[Local magnetisation in steady-states for two initial states for a variety of coupling constants $V$ : $\left|23;\downarrow\downarrow\downarrow\downarrow\downarrow\downarrow\downarrow\downarrow\downarrow\downarrow\downarrow\downarrow\downarrow\uparrow\uparrow \right>$ in both case but for two different bosonic frequencies. The total number of excitations is $M=25$.]{Local magnetisation in steady-states for two initial states for a variety of coupling constants $V$ : $\left|23;\downarrow\downarrow\downarrow\downarrow\downarrow\downarrow\downarrow\downarrow\downarrow\downarrow\downarrow\downarrow\downarrow\uparrow\uparrow \right>$ in both case but for two different bosonic frequencies. The total number of excitations is $M=25$.}
\label{spinflip_resonance}
\end{figure}

One can now study the steady-state behaviour when a spin alternation in the initial state  is located either far or close to the bosonic mode resonance energy. Fig.\ref{spinflip_resonance} compares two identical initial spin configuration: all the spins are pointing down except the two last ones pointing up. The only difference is the frequency of the bosonic mode which is either far or near the spin alternation.

In panel (a), the resonance and the spin alternation are far enough so that the two different effects do not affect each other. All the visible phenomena have therefore already been discussed in the previous subsections, both the resonance far from $\omega=6.95$ and the initial spin configuration having low entropy prevent the system from reaching a fully plane-polarised state at coupling $V=7.5$ ($V_{sup}>7.5$). On can also notice the strong overshoot where spins at large $\epsilon_i$ go either over (for those starting in a down state) or below (for those starting in an up state) zero magnetisation. In panel (b), we observe as in all the previous cases, that the resonance almost kills the effect of the memory of the initial state in its vicinity, near $\omega=13.5$. Indeed, it becomes, even at weak coupling, hard to guess what was the initial state of the spins of indexes $13$ and $14$. Nonetheless, it remains visible that the memory of the initial state has not entirely disappeared near the resonance since the local magnetisation of spin 13 is not yet zero. 

\subsection{Excitation number and population of the bosonic mode}
\label{excitationM}

Having already conjectured that a large number of bosonic excitations could enhance the tendency of the system to reach a fully plane-polarised state at a smaller coupling, we present here different plots of the local magnetisation for a varying number of excitations $M$ in the system. Going to large number of excitations \cite{bosonhugo} has been made possible by the eigenvalue-based formalism described in the first sections which makes the complexity of the solutions controlled only by the number of spins $N$ instead of growing with the number of excitations $M$ as in a traditional Bethe anstaz approach.

\begin{figure}[h]
   
    \centering
 \begin{tabular}{cc}
 \includegraphics[width=0.55\linewidth]{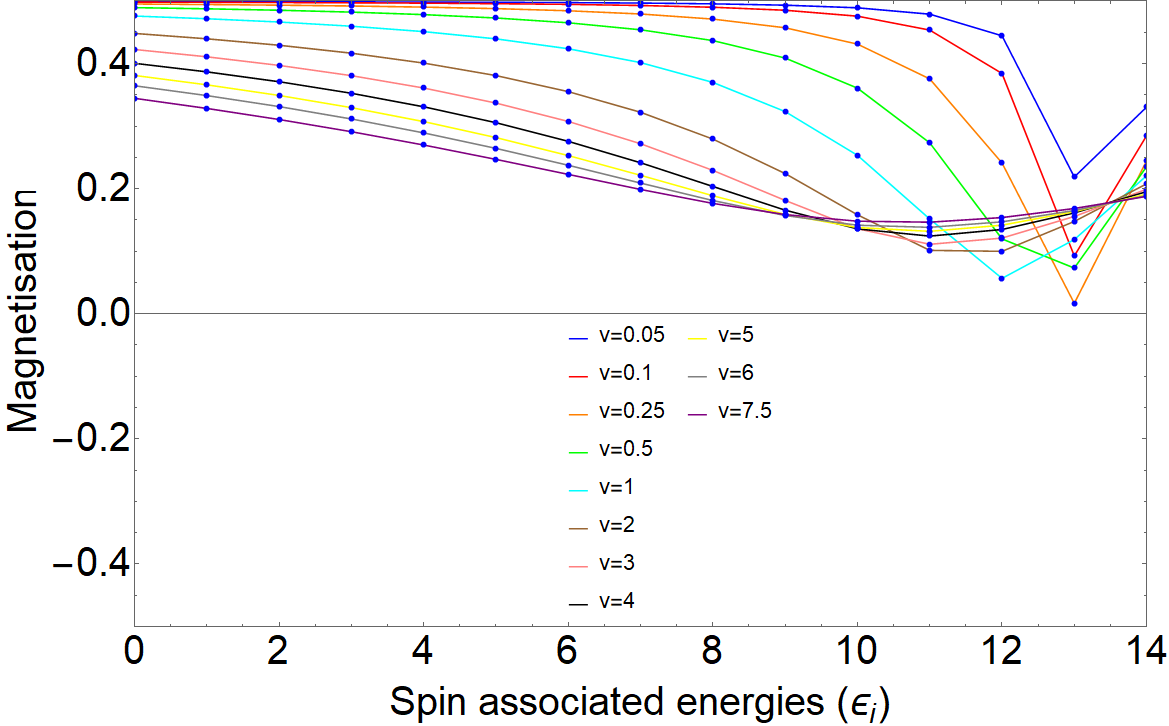}    & \includegraphics[width=0.55\linewidth]{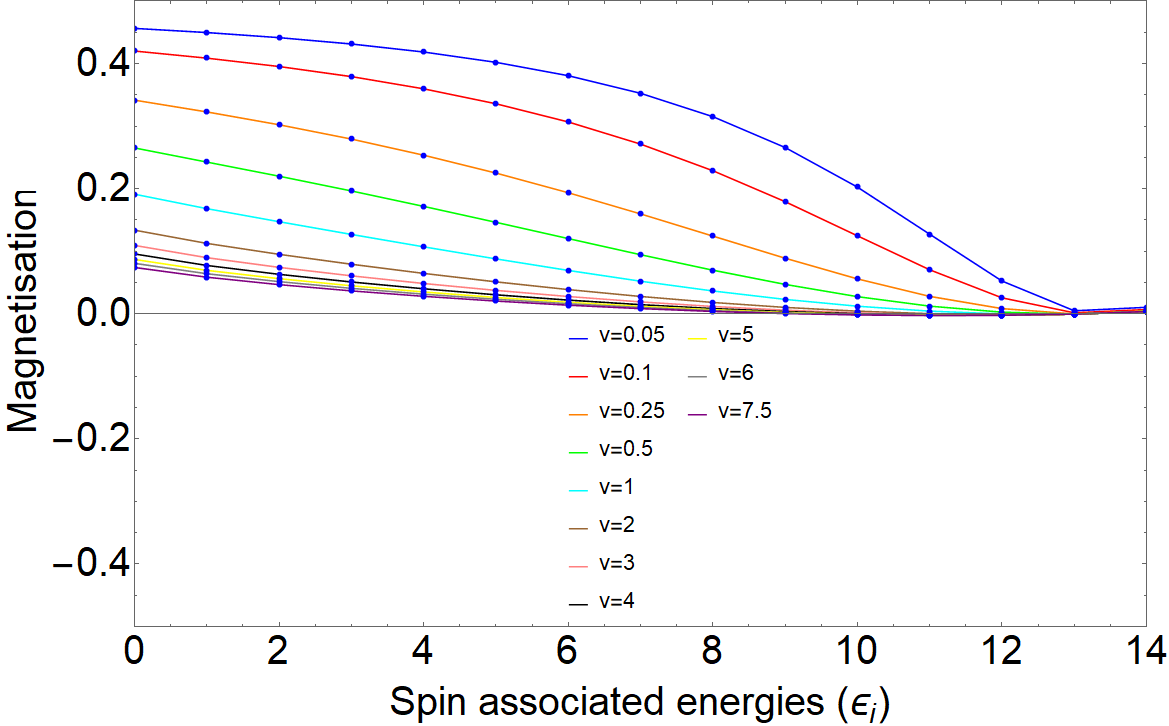} \\
 a)Initial state:$\left|0;\uparrow\uparrow\uparrow\uparrow\uparrow\uparrow\uparrow\uparrow\uparrow\uparrow\uparrow\uparrow\uparrow\uparrow\uparrow \right>$ & b) Initial state:$\left|85;\uparrow\uparrow\uparrow\uparrow\uparrow\uparrow\uparrow\uparrow\uparrow\uparrow\uparrow\uparrow\uparrow\uparrow\uparrow \right>$ \\[1em]
 \includegraphics[width=0.55\linewidth]{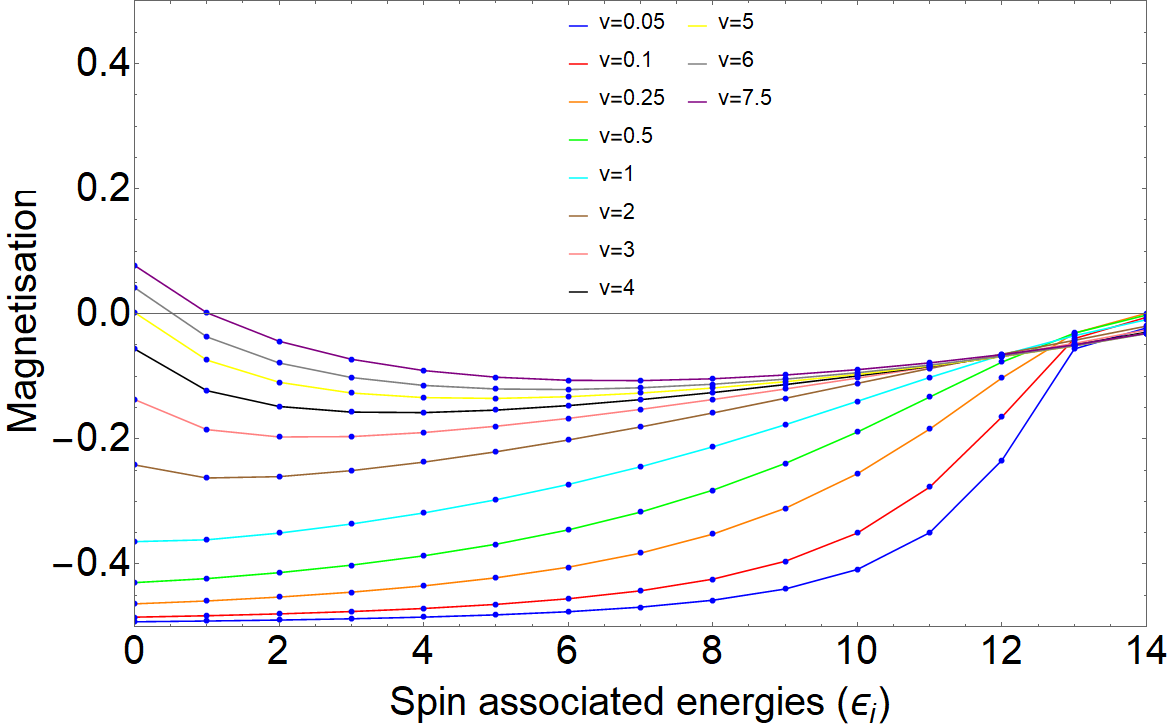}    & \includegraphics[width=0.55\linewidth]{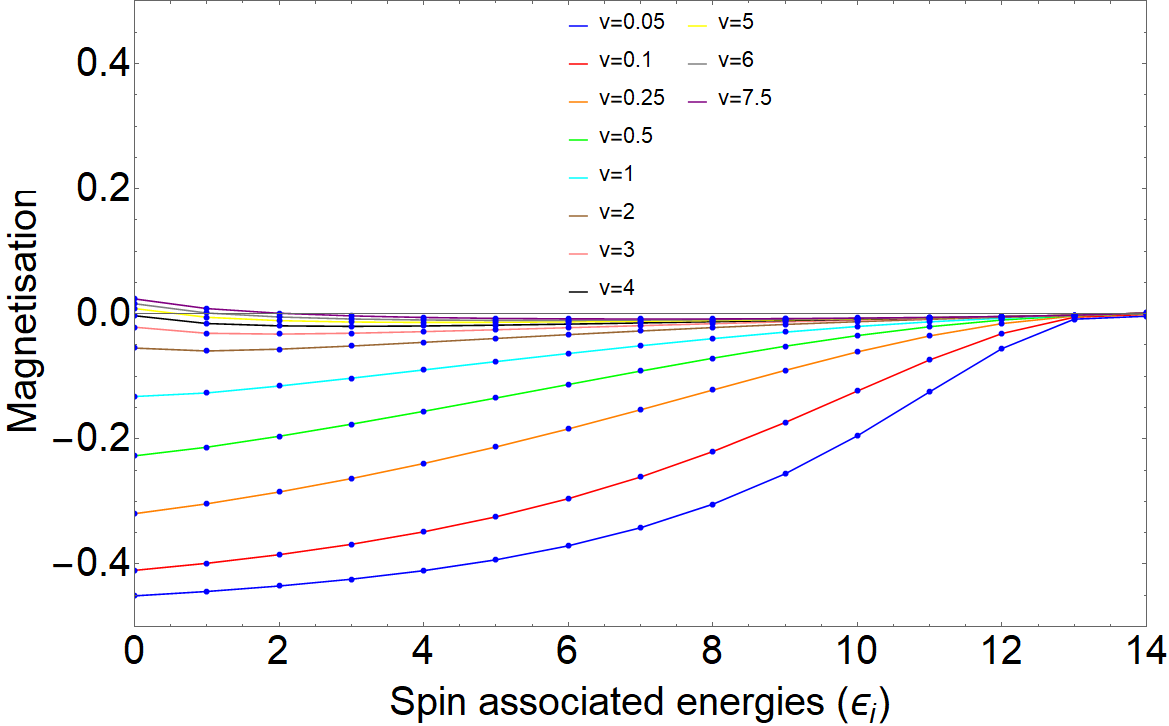} \\
 c) Initial state:$\left|15;\downarrow\downarrow\downarrow\downarrow\downarrow\downarrow\downarrow\downarrow\downarrow\downarrow\downarrow\downarrow\downarrow\downarrow\downarrow \right>$ & d) Initial state:$\left|100;\downarrow\downarrow\downarrow\downarrow\downarrow\downarrow\downarrow\downarrow\downarrow\downarrow\downarrow\downarrow\downarrow\downarrow\downarrow \right>$ \\[1em]
 \end{tabular}
 \caption{Local magnetisation in steady-states for fully up or down polarised initial states containing either $M=15$ or $M=100$ total excitations. The bosonic frequency is set to $\omega=13.5$}
 \label{fig:number}
 \end{figure}

Fig.\ref{fig:number} shows four different plots of the steady-state local magnetisation: panels (a) and (b) have a fully up initial spin state but differ by $M$ whereas panels (c) and (d) have a fully down initial polarisation but also differ by $M$. We will compare (a) with (b) then (c) with (d) to confirm our assumption on the impact of having a lot of bosonic excitations available. Knowing that without a bosonic bath, the initial configurations fully up and fully down are equivalent, we can compare (a) with (c) then (b) with (d) to see if this behaviour holds in the presence of the bosonic mode.

The comparison of (a) with (b) and (c) with (d) confirms that additional initial bosonic excitations tend to make a given coupling more efficient in approaching the zero-magnetisation steady-state, i.e. $V_{sup}$ becomes smaller for large $M$. Indeed,  for any coupling, the magnetisation of each of the spins is closer to zero for (b) and (d) than for (a) and (c). As can be seen in panel a), the complete absence of bosonic excitations in the initial state limits the efficiency of the resonance to redistribute excitations which leads to spins $12$, $13$ and $14$ maintaining a larger magnetisation despite being nearly resonant.

For a system built exclusively out of spins-$\frac{1}{2}$ with isotropic couplings, going from a fully up to a fully down-polarised state would simply amount to a change of quantisation axis and should consequently lead to the same behaviour. However, comparing (a) with (c) then (b) with (d), wich both contain the same total number of excitations, demonstrates that the nature, spin or boson, of the initial excitations actually affects the steady-state which is reached by the system. This dependency in the nature of the excitations arrises despite the fact that the coupling terms $b^\dag S^-_i$ and $b S^+_i$ are such that the process of creating (absorbing) a boson through a down (up) spin-flip are both controlled by the same single parameter $V$. Indeed, it is particularly clear in comparing (b) with (d) that the system possesses a preferred direction of magnetisation in that spins far from the resonance (of indexes $0$ and $1$) will become up-polarised in the $V = 6, V=7.5$ steady-states irrespective of their up or down initial state. While in \cite{bosonhugo} it was shown that a Bethe ansatz can be built by creating excitations on the true pseudo-vacuum $\left| 0; \downarrow \downarrow  \dots \downarrow\right>$ or by removing them from $\left| M; \uparrow \uparrow  \dots \uparrow\right>$, the second approach differs since $\left| M; \uparrow \uparrow  \dots \uparrow\right>$ is not a proper vacuum (i.e. not an eigenstate of $\mathrm{S}^2(u)$). This asymmetry leads to the overshoot, at strong coupling, for initial down polarised spins and the "reluctance" of the up polarised spins to go to a zero polarisation.


\begin{figure}[h]\hspace{1cm}\includegraphics[scale=0.4]{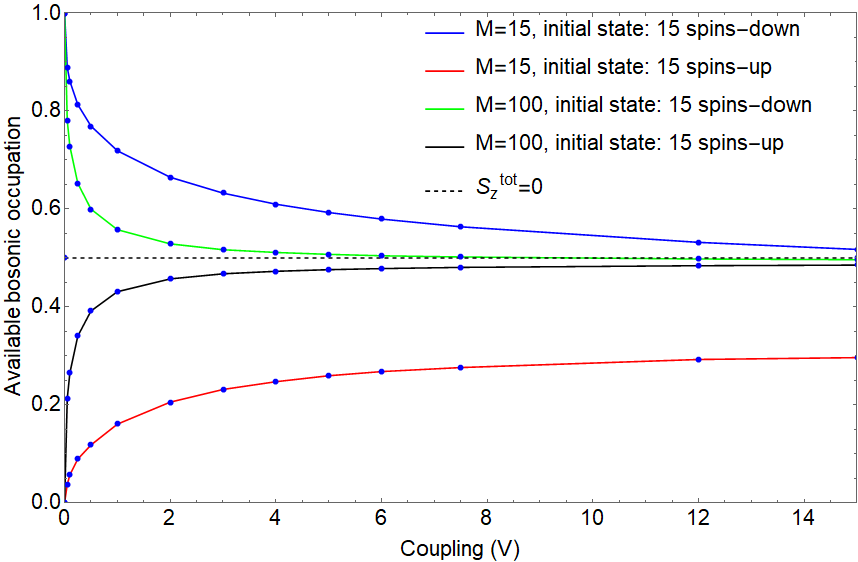}  
\vspace{-0.0cm}
\caption[ Available bosonic occupation (divided by the number of spins) in the steady-state for various initial states as a function of $V$ with the bosonic frequency set to $\omega=13.5$]{Available bosonic occupation (divided by the number of spins) in the steady-state for various initial states:$\left|0;\uparrow\uparrow\uparrow\uparrow\uparrow\uparrow\uparrow\uparrow\uparrow\uparrow\uparrow\uparrow\uparrow\uparrow\uparrow \right>$, $\left|85;\uparrow\uparrow\uparrow\uparrow\uparrow\uparrow\uparrow\uparrow\uparrow\uparrow\uparrow\uparrow\uparrow\uparrow\uparrow \right>$, $\left|15;\downarrow\downarrow\downarrow\downarrow\downarrow\downarrow\downarrow\downarrow\downarrow\downarrow\downarrow\downarrow\downarrow\downarrow\downarrow \right>$ and $\left|100;\downarrow\downarrow\downarrow\downarrow\downarrow\downarrow\downarrow\downarrow\downarrow\downarrow\downarrow\downarrow\downarrow\downarrow\downarrow \right>$ as a function of $V$. The bosonic frequency is set to $\omega=13.5$.}
\label{boson_excitation_number}
\end{figure}

The four curves of Fig.\ref{boson_excitation_number} represent the available bosonic occupation (normalised by the number of spins) in terms of $V$ in the four cases presented before. The finite number of spins N means that, for $M>N$ a minimal number of bosons $M-N$ will always be present since the system can only accommodate $N$ spin excitations. While for $M=N=15$, we plot $b^{\dag}b/N$, for $M=100$ we first remove $M-N$ from the true bosonic occupation and plot the fraction of available bosons $\left(b^{\dag}b-(M-N)\right)/N$ which are present in the steady-state. 

The global behaviour of the system, studied through the evolution of the number of bosons, confirms the picture which was described. The black dashed line, representing zero total magnetisation, is approached at weaker $V$ when the initial spin configuration is down polarised. The process of using a boson to flip a spin up is more efficient than the reverse process of flipping a spin down to create an additional boson, ultimately favouring up-polarisation of the spins. It also clearly confirms that a large number of bosons (green and black curves) enhances the redistribution of the excitations and thus favours the emergence of a zero-magnetisation steady-state at weaker couplings, in other words $V_{sup}$ is smaller for systems initially containing an large number of bosonic excitations.

\subsection{Individual spin behaviour }

Through the various numerical results concerning the steady-state of spin-boson Richardson-Gaudin models, a qualitative description of the various observed features has emerged. At very strong couplings, independently of the details of the initial state or the model's parameter, the steady-state which is realised seems universally defined by zero average magnetisation for each of the spins. However, for weaker couplings $V$, the steady-state appears as an interrupted (pseudo-)relaxation towards this universal state, $V$ therefore playing a role analogous to time in a true relaxation process.

The efficiency of this process, namely how strong $V$ has to be for a given spin to reach a zero-magnetisation steady-state, if affected by a number of global and local factors. First, highly entropic initial states favours faster (in $V$) relaxation, a fact which is true globally as well as locally since spin alternations make the concerned spins relax more efficiently. Spin whose "energy gap" $\epsilon_i$ is nearly resonant with the value of the bosonic frequency $\omega$ are also relaxing more efficiently, which has for consequence that $\omega$ in the middle of the energy band leads to faster global relaxation since it minimises the width of the $\omega-\epsilon_i$ band. Moreover, a large number of bosonic excitations in the initial state also mean that the zero-magnetisation steady-state is reached for smaller values of the coupling $V$.

Explicitly plotting the steady-state local magnetisation of four spins as a function of the coupling $V$, we show that the  "evolution" of individual spins as a function of $V$ do not clearly lead to a universal approach to the strong coupling zero-magnetisation steady-state.

\begin{figure}[h!]\hspace{1cm}\includegraphics[scale=0.4]{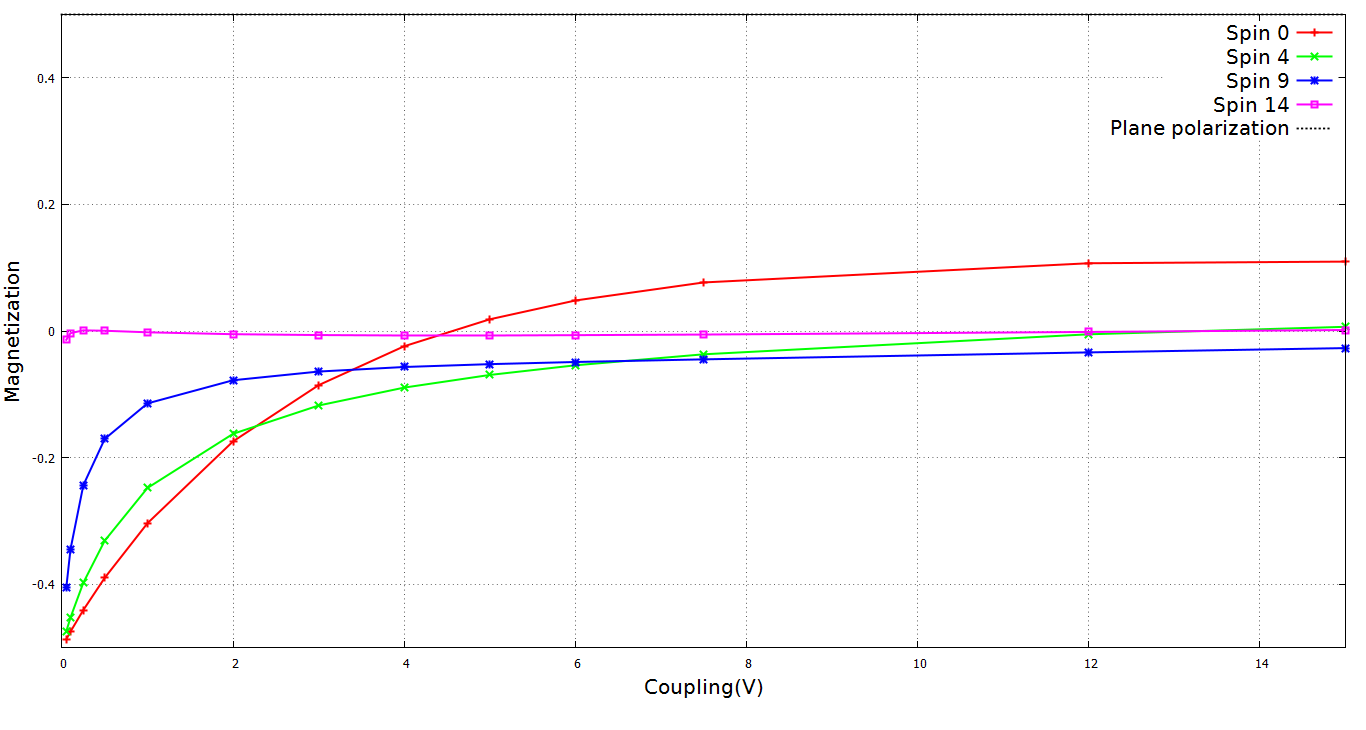}  
\vspace{-0.5cm}
\caption{Local magnetisation of four spins, associated with $\epsilon_{0},\epsilon_{4},\epsilon_{9}$ and $\epsilon_{14}$ when $V$ goes from $0$ to $15$ with $M=25$ and $\omega=13.5$.  Initial state: $\left|25;\downarrow\downarrow\downarrow\downarrow\downarrow\downarrow\downarrow\downarrow\downarrow\downarrow\downarrow\downarrow\downarrow\downarrow\downarrow \right>$ with a bosonic frequency set at $\omega=13.5$} 
\label{complexity}
\end{figure}

Fig.\ref{complexity} shows the evolution, as a function of $V$, of the four spins, with associated energies $\epsilon_0$, $\epsilon_4$, $\epsilon_9$ and $\epsilon_{14}$, starting from the following initial state $\left|25;\downarrow\downarrow\downarrow\downarrow\downarrow\downarrow\downarrow\downarrow\downarrow\downarrow\downarrow\downarrow\downarrow\downarrow\downarrow \right>$ with a bosonic frequency set at $\omega=13.5$. Although the features described in this work are all consistent with the observed qualitative behaviour, the individual spins do not have a similar V-behaviour. In fact, no single universal form for the "decay": exponential, power law, etc., can properly fit the four curves presented here. However, the resonance effect is clearly visible: the spin which are the closest to resonance with $\omega$ (14 and 9) decay to zero-magnetisation rapidly, while spins far from resonance (0 and 4) require much higher values of $V$ to do so. They also overshoot the zero-magnetisation (especially spin 0) which could be described as resulting from underdamping, but a quantitative description  through the definition of some relaxation model (in V) seems to go beyond what is feasible in these systems.

\section{Conclusions}

Using the theoretical background developed in the first sections of this work, namely the rewriting of the required physical quantities including form factors, in terms of the eigenvalue-based variables $\Lambda(\epsilon_{i})$, we have managed to study efficiently the steady-state properties of spin-boson Richardson-Gaudin integrable models even when a large number of excitations is present.

These results which describe the long-time average local physical quantities calculated through the diagonal ensemble are universal for every Hamiltonian of this class, some of which having radically different structures in the couplings involved. A qualitative description of these numerical results has also emerged. Certain aspects of the coupling dependence of the steady-state are naturally understood in a Tavis-Cummings Hamiltonian since it can be understood as a resonant process which leads to very efficient relaxation of resonant spins whose magnetisation disappears in the steady-state even at very weak coupling. However this resonance remains in central-spin-like systems for which such a description would be much less natural.

Globally, it appears that, at large enough couplings, the steady-state reached by arbitrary initial states is characterised by zero average magnetisation of each individual spin. Highly entropic initial spin configurations, a bosonic frequency in the middle of the energy band therefore maximising the impact of the resonant-effect, as well as a large number of bosonic excitations all appear to favour the appearance of this universal zero-magnetisation steady-state at weaker coupling.


\section*{References} 

\bibliographystyle{unsrt}
\bibliography{bibliography}

\end{document}